\documentclass[aps,twocolumn,prd,reprint,nofootinbib]{revtex4-1}

\usepackage{amsmath,amssymb,amsfonts}
\usepackage{bbm}
\usepackage{graphicx}
\usepackage[T1]{fontenc}
\usepackage[utf8]{inputenc}
\usepackage[dvipsnames]{xcolor}
\definecolor{bluecite}{HTML}{0875b7}
\usepackage[colorlinks=true,urlcolor=Emerald,linkcolor=bluecite,citecolor=bluecite]{hyperref}
\usepackage{acronym}
\usepackage[normalem]{ulem}
\usepackage{orcidlink}

\newcommand{\dimlessp}{\ensuremath{\tilde p}}
\newcommand{\dimlesspsq}{\ensuremath{\dimlessp^2}}
\newcommand{\physdimlesspsq}{\ensuremath{X}} 

\newacro{RG}[RG]{renormalisation group}

\makeatletter

\renewcommand\onecolumngrid{%
  \do@columngrid{one}{\@ne}%
  \def\set@footnotewidth{\onecolumngrid}%
  \def\footnoterule{\kern-6pt\hrule width 1.5in\kern6pt}%
}

\renewcommand\twocolumngrid{%
  \def\footnoterule{%
    \dimen@\skip\footins\divide\dimen@\thr@@
    \kern-\dimen@\hrule width.5in\kern\dimen@}%
  \do@columngrid{mlt}{\tw@}%
}

\makeatother

\allowdisplaybreaks

\begin{document}

\title{The pole truth: an analytical graviton propagator from Asymptotic Safety}

\author{Benjamin Knorr\,\orcidlink{0000-0001-6700-6501}\,}
\email[]{knorr@thphys.uni-heidelberg.de}
\affiliation{Institut f\"ur Theoretische Physik, Universit\"at Heidelberg, Philosophenweg 12, 69120 Heidelberg, Germany}

\date{\today}

\begin{abstract}
We derive an analytical approximation for the graviton propagator from Asymptotic Safety. We find neither extra poles nor indications of unitarity or causality violations in the spin-two sector. Our results strengthen the case that Asymptotic Safety does not introduce new degrees of freedom, and thus propagates the same field content as General Relativity. We also identify the underlying mechanism: the residues of spurious poles in finite-order derivative expansions approach zero as the order is increased.
\end{abstract}
\maketitle

\textit{Introduction.}---Formulating a consistent and phenomenologically viable theory of quantum gravity is one of the major open problems in theoretical physics. Critical consistency requirements are unitarity and causality; different approaches put varying amounts of weight on these aspects~\cite{Basile:2024oms, Buoninfante:2024yth}. Ultimately, this is a computational question: given specific assumptions, is the theory consistent? In this Letter, we strengthen existing evidence~\cite{Bonanno:2021squ, Fehre:2021eob, Pawlowski:2025etp} that one specific approach, Asymptotic Safety~\cite{Percacci:2017fkn, Reuter:2019byg, Knorr:2022dsx, Eichhorn:2022gku, Morris:2022btf, Martini:2022sll, Wetterich:2022ncl, Platania:2023srt, Saueressig:2023irs, Pawlowski:2023gym, Bonanno:2024xne}, is unitary, causal, and does not propagate extra modes beyond a standard massless graviton.

Asymptotic Safety is based on the idea of an interacting \ac{RG} fixed point that tames the theory at high energies entirely within the framework of quantum field theory. By now, the existence of such a fixed point has been established convincingly, for reviews see~\cite{Percacci:2017fkn, Reuter:2019byg, Knorr:2022dsx, Eichhorn:2022gku, Morris:2022btf, Martini:2022sll, Wetterich:2022ncl, Platania:2023srt, Saueressig:2023irs, Pawlowski:2023gym, Bonanno:2024xne}. Nevertheless, it is considered an important open question whether the theory is unitary and causal~\cite{Bonanno:2020bil}. Recent work computed the spin-two part of the momentum-dependent Lorentzian graviton propagator numerically~\cite{Bonanno:2021squ, Fehre:2021eob, Pawlowski:2025etp},\footnote{See also~\cite{Glaviano:2026lew, Glaviano:2026aoe} for recent results in a proper-time formulation.} finding evidence that there is only a massless pole, and that the spectral function is positive. No indications for massive ghosts have been found, which appear only as truncation artefacts in derivative expansions~\cite{Platania:2020knd, Platania:2022gtt}.

In this work, we put forward an \emph{analytical} understanding of the graviton propagator in Asymptotic Safety, following the ideas of~\cite{Knorr:2026vax}. We resolve all degrees of freedom, and confirm previous numerical results where available~\cite{Bonanno:2021squ, Fehre:2021eob, Pawlowski:2025etp}. Within our approximation, we show analytically that the propagator is free from additional poles, and has the complex structure expected from a unitary and causal theory. We also identify the mechanism that avoids spurious poles that appear in finite-order derivative expansions.

Our results can be used as an input for future computations of graviton correlation functions and gravitational scattering amplitudes~\cite{Draper:2020bop, Draper:2020knh, Knorr:2022lzn, Pastor-Gutierrez:2024sbt, Kher:2025rve, Knorr:2026vax, Chiesa:2026tlz}, but also to derive analytical quantum equations of motion containing all terms up to second order in curvature. This opens up investigations of quantum black holes from first principles~\cite{Pawlowski:2023dda}.

\textit{Graviton propagator.}---In a gauge-fixed setting, the graviton propagator can be decomposed into its irreducible representations: the physical spin-two part, related to the two polarisations of gravitational waves, a spin-one and a spin-zero gauge mode, and another spin-zero mode which is related to the conformal factor. The latter propagates off-shell but not on-shell, so it contributes to loop diagrams. Via the Faddeev-Popov procedure, there are also spin-one and spin-zero ghosts which cancel the gauge modes of the graviton. The ghost propagator is discussed in the Supplemental Material.

We parameterise the graviton propagator in flat spacetime via form factors~\cite{Knorr:2019atm}. The relevant part of the effective action reads
\begin{equation}\label{eq:effaction}
\begin{aligned}
    \Gamma = \frac{1}{16\pi G_N} \int \text{d}^4x \, \sqrt{-g} \Bigg[ &R + \frac{1}{6} R F_R(\Box) R \\
    &- \frac{1}{2} C^{\mu\nu\rho\sigma} F_C(\Box) C_{\mu\nu\rho\sigma} \Bigg] \, .
\end{aligned}
\end{equation}
In this, $C$ is the Weyl tensor, $R$ is the Ricci scalar, $G_N$ is Newton's constant, and $F_{R,C}$ are form factors depending on the covariant d'Alembertian $\Box=-D^2$. In this basis, the form factor $F_C$ ($F_R$) only contributes to the spin-two (zero) part of the graviton propagator $\mathcal G^{\mu\nu\rho\sigma}$ in Minkowski space. For a graviton with momentum $p$, the latter reads
\begin{equation}\label{eq:gravprop}
\begin{aligned}
    \mathcal G^{\mu\nu\rho\sigma}(p^2) &\propto G_2(p^2) \Pi_2^{\mu\nu\rho\sigma} + G_0(p^2) \Pi_0^{\mu\nu\rho\sigma} + \dots \, , \\
    G_2(p^2) &= \frac{G_N}{p^2(1+p^2 F_C(p^2))} \, , \\
    G_0(p^2) &= -\frac{G_N}{p^2(1+p^2 F_R(p^2))} \, ,
\end{aligned}
\end{equation}
where $G_{2,0}$ and $\Pi_{2,0}^{\mu\nu\rho\sigma}$ are the propagator functions and projectors of the spin-two and zero sectors~\cite{Knorr:2021niv}, respectively, and the dots indicate the gauge modes. The form factors $F_{R,C}$ are thus in one-to-one correspondence to the momentum-dependent graviton propagator in Minkowski space.

For physical states, unitarity precludes poles with negative residues in the propagator. Causality constrains the propagator to be analytic in the upper half-plane in energy, with singularities restricted to physical poles and timelike branch cuts. We provide analytical but non-perturbative approximations for $G_{2,0}$, and thereby evidence for unitarity and causality in Asymptotic Safety. 

For fields creating physical states, the time-ordered propagator admits a K\"all\'en-Lehmann spectral representation in terms of the spectral function $\rho$~\cite{Kallen:1952zz, Lehmann:1954xi},
\begin{equation}
    G(p_0) =\mathbf{i} \int_0^\infty \frac{\text{d}\lambda}{\pi} \frac{\lambda \, \rho(\lambda)}{p_0^2 - \lambda^2 + \mathbf{i}\epsilon} \, .
\end{equation}
For such states, $\rho$ is positive and can be normalised,
\begin{equation}
    \int_0^\infty \frac{\text{d}\lambda}{\pi} \lambda \, \rho(\lambda) = 1 \, .
\end{equation}
Gauge-fixed fields like the graviton are not directly linked to gauge-invariant asymptotic states, and thus do not necessarily admit a (positive) spectral representation. In agreement with earlier results~\cite{Bonanno:2021squ}, we find that all modes in fact admit such a representation, but with a non-normalisable spectral function.

\textit{Correlation functions in Asymptotic Safety.}---Our setup is based on the well-developed framework of momentum-dependent correlation functions~\cite{Christiansen:2012rx, Christiansen:2014raa, Christiansen:2015rva, Denz:2016qks}, for reviews see~\cite{Pawlowski:2020qer, Pawlowski:2023gym}. The functional \ac{RG}~\cite{Wetterich:1992yh, Reuter:1996cp, Dupuis:2020fhh} provides a coupled tower of functional differential equations for these correlation functions, which is closed by a truncation. The current state of the art takes a seed action to model the higher-order vertices that are not resolved independently. In our case, we resolve the complete two-point function, and use the Einstein-Hilbert action for the three- and four-graviton vertex, using the results of~\cite{Knorr:2021niv}. Evidence for the apparent convergence of this scheme was provided, e.g., in~\cite{Christiansen:2015rva, Denz:2016qks}. It is crucial to note that using such a seed action for the vertices does not imply the absence of higher-order terms in our results. Rather, as in perturbation theory, higher-order terms are generated through loop effects. Thus, using the Einstein-Hilbert action does not entail that the propagator automatically has a tree-level form.

We model the \ac{RG} flow of the gravitational coupling via
\begin{equation}
    g_k = \frac{G_N k^2}{1+\frac{G_N k^2}{g_\ast}} \, .
\end{equation}
In this, $k$ is the \ac{RG} scale connected to the infrared regulator of the functional \ac{RG}, $G_N$ is the physical Newton's coupling, $g_k$ is its dimensionless, $k$-running version, and $g_\ast>0$ is its fixed-point value. This model~\cite{Bonanno:2000ep} was shown to be accurate in~\cite{Fehre:2021eob}. Further technical choices are provided in the Supplemental Material.

With this, we arrive at a set of four linear, coupled integral equations for the momentum-dependent anomalous dimensions of the spin-two and spin-zero part of the graviton as well as the spin-one and spin-zero components of the Faddeev-Popov ghost~\cite{Knorr:2021niv}. We furthermore simplify the setup by neglecting the feedback of the anomalous dimensions in diagrams. In our gauge choice, this is an excellent approximation, see the Supplemental Material.

The anomalous dimensions are then integrated to obtain the wave function renormalisation factors $Z$, which are the non-trivial dressing factors of the propagators. For example, for the spin-two propagator, we have
\begin{equation}
    G_2(p^2) = \frac{G_N}{p^2 Z_2(p^2)} \quad \Rightarrow \quad F_C(p^2) = \frac{Z_2(p^2)-1}{p^2} \, .
\end{equation}

\textit{Results.}---We focus our discussion on the Landau gauge with gauge parameter $\beta=-1$. Results for arbitrary $\beta$ are reported in the Supplemental Material. For the fixed-point value of the gravitational coupling, we chose $g_\ast=1/2$, which lies in the reliable regime of our approximation of neglecting the self-feedback of the anomalous dimensions.

\begin{figure}[!t]
\centering
\includegraphics[width=\linewidth]{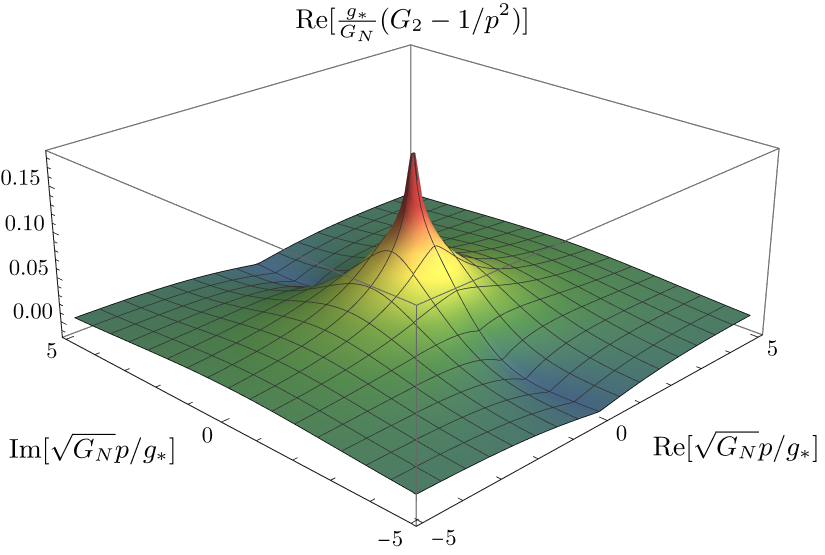} \\ \includegraphics[width=\linewidth]{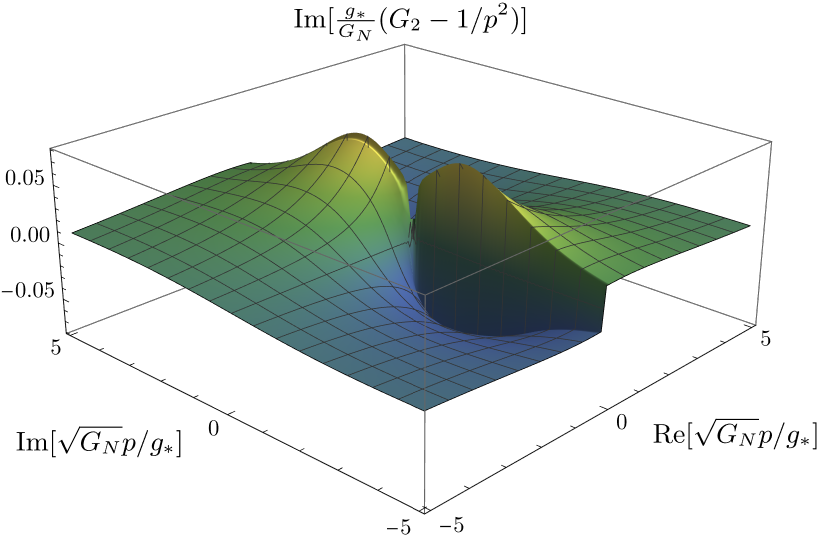} 
\caption{\label{fig:G2}Real and imaginary part of the spin-two propagator function $G_2$ in the complex plane in Planck units, as a function of momentum in Planck units. The massless pole at $p^2=0$ has been subtracted, and we have used $g_\ast=1/2$. The imaginary part displays a timelike branch cut that gives rise to a positive spectral function.}
\end{figure}

\begin{figure}[!t]
\centering
\includegraphics[width=0.9\linewidth]{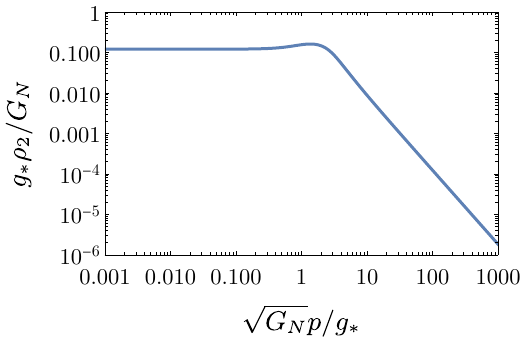}
\caption{\label{fig:rho2}Spectral function $\rho_2$ of the spin-two part of the graviton in Planck units, as a function of momentum in Planck units. It is positive on the entire positive real axis. We have used $g_\ast=1/2$.}
\end{figure}

In \autoref{fig:G2}, we present the real and imaginary parts of the spin-two propagator $G_2(p^2)$ in the complex plane in Planck units, where we have subtracted the pole at $p^2=0$. The qualitative picture confirms previous numerical results obtained from Euclidean~\cite{Bonanno:2021squ} and Lorentzian~\cite{Fehre:2021eob, Pawlowski:2025etp} computations. The imaginary part admits a branch cut along the timelike axis, giving rise to a positive spectral function shown in \autoref{fig:rho2}. The latter is not normalisable. Since a linear graviton is not a gauge-invariant asymptotic state, this is a priori not a problem, as discussed in detail in~\cite{Bonanno:2021squ}.

On top of the qualitative agreement, there is also quantitative agreement with~\cite{Bonanno:2021squ}, in the following sense. For the comparison, we switch to $\beta=1$ used in~\cite{Bonanno:2021squ}, and expand the spin-two propagator for small $p^2$,
\begin{equation}
    G_N^{-1}G_2(p^2) \sim \frac{1}{p^2} - A_h \ln G_N p^2 + C_h \, .
\end{equation}
The coefficient $A_h$ is regulator-independent but gauge-dependent. For $\beta=1$, it reads
\begin{equation}
    A_h(\beta=1) = \frac{7}{20\pi} \, .
\end{equation}
The expression for general $\beta$ can be extracted from the expansion of the form factors, \eqref{eq:wc}. The Wilson coefficient $C_h$ in our computation depends on the fixed-point value $g_\ast$. If we use the value provided in~\cite{Bonanno:2021squ}, $g_\ast=570\pi/833$, we obtain
\begin{equation}
    C_h = \frac{5927+3360(\ln 2g_\ast -\gamma)}{9600\pi} \approx 0.2947 \, .
\end{equation}
This agrees with the result $C_h\approx 0.29$ given in~\cite{Bonanno:2021squ} to all significant digits. This is remarkable, because at least in truncations, Wilson coefficients are regulator-dependent, and an entirely different regulator was chosen in~\cite{Bonanno:2021squ}. The agreement suggests that both our results and~\cite{Bonanno:2021squ} are quantitatively reliable.

For small momenta, the spin-two spectral function is positive only in a finite range of gauge parameters,
\begin{equation}
    -2.19 < \beta < 1.72 \, || \, 1.98 < \beta < 3 \, .
\end{equation}
The upper bound $\beta<3$ is due to the gauge fixing being incomplete~\cite{Gies:2015tca}. Consequently, for gauge-dependent quantities, the regime near $\beta=3$ is less reliable. The range is in one-to-one correspondence with the sign of $A_h$, and thus a regulator-independent statement. This re-emphasises that a linear graviton is not a gauge-invariant asymptotic state, and is reminiscent of the behaviour of the gluon~\cite{Cyrol:2018xeq}.

\begin{figure}[!t]
\centering
\includegraphics[width=\linewidth]{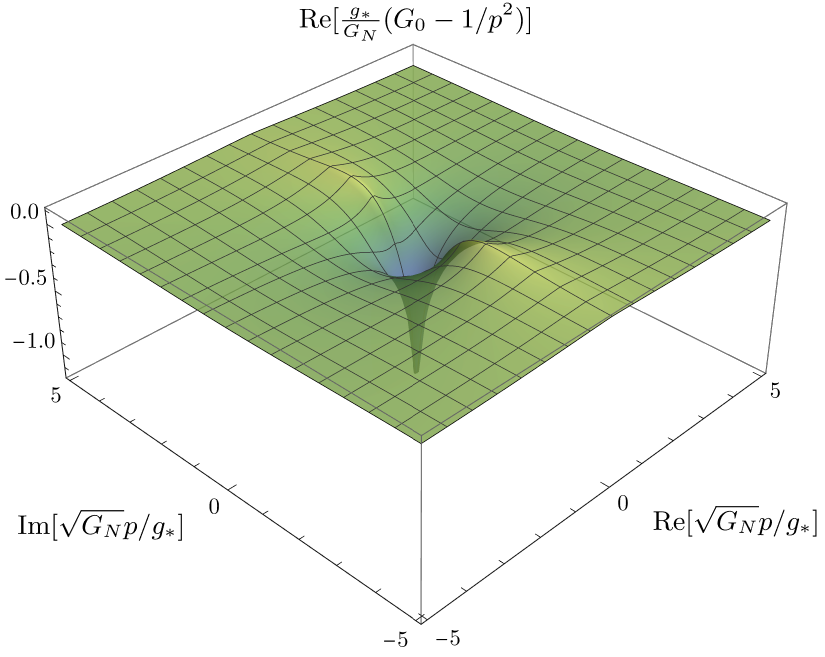} \\
\includegraphics[width=\linewidth]{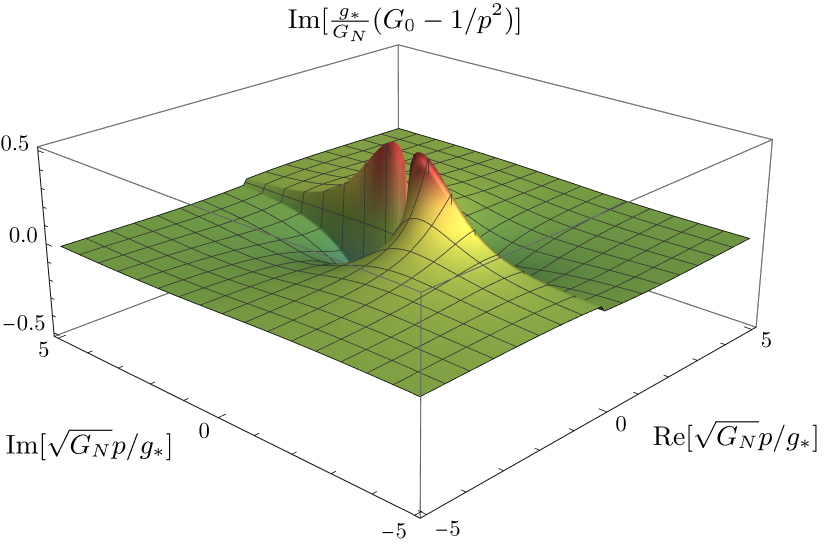} 
\caption{\label{fig:G0}Real and imaginary part of the spin-zero propagator function $G_0$ in the complex plane in Planck units, as a function of momentum in Planck units. The massless pole at $p^2=0$ has been subtracted, and we have used $g_\ast=1/2$. The imaginary part displays a timelike branch cut that gives rise to a negative spectral function.}
\end{figure}

For large momenta, the spin-two propagator falls off as dictated by its fixed-point anomalous dimension (given in \eqref{eq:eta2}),
\begin{equation}
    G_N^{-1}G_2(p^2) \sim \frac{c_\beta^2}{p^2} \left( \frac{G_N p^2}{g_\ast} \right)^{\frac{\eta_{2\ast}(0)}{2}} \, ,
\end{equation}
where $c_\beta^2>0$ is a gauge- and $g_\ast$-dependent constant.

\begin{figure}[!t]
\centering
\includegraphics[width=0.9\linewidth]{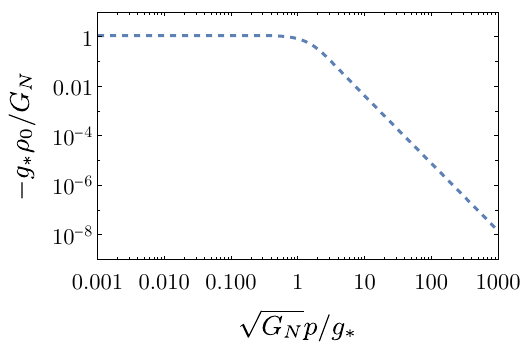}
\caption{\label{fig:rho0}Spectral function $\rho_0$ of the spin-zero part of the graviton in Planck units, as a function of momentum in Planck units. It is negative on the entire positive real axis, indicated by the dashing. We have used $g_\ast=1/2$.}
\end{figure}

The spin-zero propagator is shown in \autoref{fig:G0}, and its spectral function is depicted in \autoref{fig:rho0}. Its behaviour is the sign-reversed analogue of the spin-two mode, as expected from our convention  \eqref{eq:gravprop}. The negative spectral function reflects the familiar wrong sign of the off-shell conformal factor; since the quantum-corrected spin-zero propagator has no additional poles, this sector is not promoted to a new propagating scalar degree of freedom. This strengthens the evidence that Asymptotic Safety does not introduce extra propagating modes beyond those of General Relativity.

Other than the overall sign difference, the qualitative behaviour of the two modes is the same in many regards. For small momenta, we again find
\begin{equation}
    G_N^{-1}G_0(p^2) \sim -\left( \frac{1}{p^2} - \tilde A_h \ln G_N p^2 + \tilde C_h \right) \, .
\end{equation}
The expressions for $\tilde A_h, \tilde C_h$ can be extracted from \eqref{eq:wc}. The spin-zero spectral function at vanishing argument is always negative, independently of $\beta$.

A difference between the two modes appears for large momenta. The spin-zero mode is not momentum-local~\cite{Christiansen:2015rva}, for a detailed discussion see~\cite{Pawlowski:2020qer}. Consequently, the fall-off of $G_0$ is determined by a \emph{difference} of the fixed-point spin-zero anomalous dimension~\cite{Knorr:2021niv} (given in \eqref{eq:eta0}),
\begin{equation}
    G_N^{-1}G_0(p^2) \sim -\frac{c_\beta^0}{p^2} \left( \frac{G_N p^2}{g_\ast} \right)^{\frac{\eta_{0\ast}(0)-\eta_{0\ast}(\infty)}{2}} \, ,
\end{equation}
where once again $c_\beta^0>0$ is gauge-dependent.

\textit{Unitarity mechanism.}---In a derivative expansion of the effective action~\eqref{eq:effaction} of increasing but finite order (that is, a Taylor expansion of the form factors $F_{R,C}$), the propagator features an increasing number of poles, as evident from~\eqref{eq:gravprop}. Some of them are necessarily ghostly, similar to those of quadratic gravity~\cite{Stelle:1976gc, Stelle:1977ry}. As we see from the fully momentum-dependent propagator, see also~\cite{Bonanno:2021squ, Fehre:2021eob, Pawlowski:2025etp}, such spurious poles are truncation artefacts. It is thus worthwhile to identify the mechanism that prevents the propagator from having additional poles in our results. Different such mechanisms have been proposed in~\cite{Platania:2020knd, Platania:2022gtt}.

\begin{figure}[!t]
\centering
\includegraphics[width=0.9\linewidth]{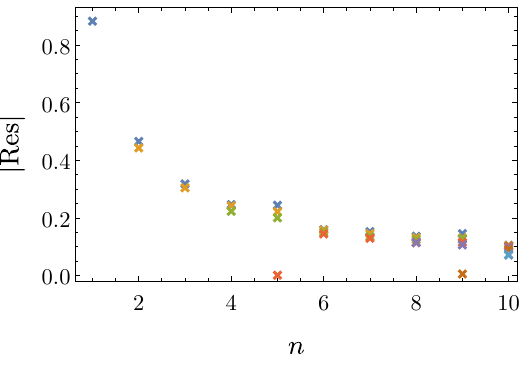}
\caption{\label{fig:spur_res}Absolute values of residues of spurious poles of the spin-two propagator in Planck units, when expanding $Z_2$ in small $p^2$ up to order $n$. As $n$ increases, the residues tend to zero. We only considered zeros with $|G_N p^2|\leq20$. Colours refer to an ordering in absolute value of the residue only, and do not identify poles across different truncation orders.}
\end{figure}

For this, we expand $Z_2$ in small $p^2$ and look for non-trivial zeros. Due to the branch cut, this is not a Taylor expansion, but it also features logarithms, see \eqref{eq:wc}. We employ the argument principle to determine the number of zeros of $Z_2$ with $|G_N p^2|\leq20$, again with $\beta=-1$ and $g_\ast=1/2$.\footnote{More zeros can appear at significantly larger values, but they are not reliable in an expansion in small $p^2$. For example, keeping the first two non-trivial terms, we find an additional zero at $G_N p^2 \simeq 10^{26}$.}

Within this region, and in an expansion up to $p^{20}$, we generally find that the number of zeros never decreases with increasing order. The absolute values of the residues of the propagator at these points are driven to zero in successive orders, see \autoref{fig:spur_res}. This realises one of the mechanisms proposed in~\cite{Platania:2020knd, Platania:2022gtt}.

\textit{Complex structure at finite cutoff.}---For the well-behaved complex structure of the propagator, it is crucial that all quantum fluctuations are integrated out, that is, that the infrared regulator of the functional \ac{RG} has been taken to zero, $k\to0$. At finite cutoff, we find regulator artefacts, including infinitely many additional poles in the complex plane. This is a manifestation of the difficulty of finding a regulator that fulfils a number of desirable properties: maintaining Lorentz invariance, inducing a finite \ac{RG} flow, and respecting causality even at finite cutoff scales~\cite{Braun:2022mgx}. At present, no such regulator is known, so in practice one of these properties is sacrificed. In a spectral formulation of the functional \ac{RG}~\cite{Horak:2020eng, Horak:2021pfr, Fehre:2021eob, Braun:2022mgx}, typically a Callan-Symanzik regulator is used. It requires an additional ultraviolet regularisation, but keeps the simple pole structure intact even at finite cutoff. It is reassuring that the expected complex structure also emerges in our case when the cutoff is removed, as was also seen in previous numerical results based on a linear regulator~\cite{Bonanno:2021squ, Fehre:2021eob, Pawlowski:2025etp}.

\textit{Discussion.}---We provide analytical first-principles insights into the non-perturbative graviton propagator from Asymptotic Safety. Our results provide significant evidence that Asymptotic Safety is unitary and causal, thus addressing a critical roadblock for the theory.

Specifically, we find no additional poles beyond the massless pole. Moreover, the spectral function of the spin-two (zero) sector is positive (negative). This agrees with previous numerical computations of the spin-two sector. Taking the physical limit of vanishing infrared cutoff is crucial: at finite cutoff, unphysical artefacts appear. We also identify the mechanism that removes spurious poles in truncations: the residues of spurious poles in a derivative expansion approach zero. Our results strengthen the existing evidence that Asymptotic Safety provides a minimal ultraviolet completion for General Relativity.

Our results are \emph{analytical}. All the above properties can be shown explicitly, without relying on numerical approximations or extrapolation. This is especially useful because the Wick rotation needed to access Lorentzian momenta is technically challenging for numerical data. We have shown that our approximation is quantitatively accurate for a relevant range of the gravitational coupling.

Our analytical results can be used directly to derive and investigate quantum equations of motion in Asymptotic Safety, for example to search for quantum black hole solutions~\cite{Pawlowski:2023dda} and to study singularity resolution~\cite{Bosma:2019aiu}. They also represent a central ingredient in quantum-gravitational scattering~\cite{Draper:2020bop, Draper:2020knh, Knorr:2022lzn, Pastor-Gutierrez:2024sbt, Kher:2025rve, Knorr:2026vax, Chiesa:2026tlz}.

The approximation method presented here, and originally introduced in~\cite{Knorr:2026vax} to compute the gravitationally induced quartic scalar self-interaction, can be extended to other correlation functions. This will eventually allow us to gain analytical insights into asymptotically safe scattering even in the non-perturbative high-energy regime.

\textit{Acknowledgments.}---I would like to thank A.~Eichhorn for interesting discussions and constructive feedback on an earlier draft.

\bibliography{bib}

\begin{widetext}
\clearpage
\onecolumngrid

\appendix

\section{Anomalous dimensions}

In this appendix, we report the full expressions for the anomalous dimensions of all modes of the graviton and the corresponding Faddeev-Popov ghost in the leading-order approximation. All results are also available in a supplemental Mathematica notebook. For this, we use the results of~\cite{Knorr:2021niv}, and employ a Landau gauge with an arbitrary second gauge parameter $\beta$. The leading-order approximation consists in neglecting the anomalous dimensions in the loop terms. For convenience, we introduce
\begin{equation}
    \tilde\beta = \frac{2}{3-\beta} \, .
\end{equation}
We furthermore identify all avatars of Newton's coupling by a single coupling $g$. The regulator shape is given by
\begin{equation}
    R_k(z) = \frac{z}{e^z-1} \, .
\end{equation}
Below, all momenta are measured in units of the functional \ac{RG} scale $k$,
\begin{equation}
    \dimlesspsq = p^2/k^2 \, .
\end{equation}
Within this setup, the anomalous dimensions $\eta_{2,0,c^T,c^L}$ (corresponding to spin two, spin zero, transverse ghost, and longitudinal ghost, respectively) can be parameterised as
\begin{equation}\label{eq:etagaugedep}
    \eta_2 = g \, \sum_{i=0}^4 \eta_2^{(i)} \tilde \beta^i \, , \qquad \eta_0 = g \, \sum_{i=0}^4 \eta_0^{(i)} \tilde \beta^i \, , \qquad \eta_{c^T} = g \, \sum_{i=0}^3 \eta_{c^T}^{(i)} \tilde \beta^i \, , \qquad \eta_{c^L} = g \, \sum_{i=0}^4 \eta_{c^L}^{(i)} \tilde \beta^i \, .
\end{equation}
The individual contributions are shown in \autoref{fig:eta_gaugedep}.

\begin{figure*}[ht]
\centering
\includegraphics[width=0.8\linewidth]{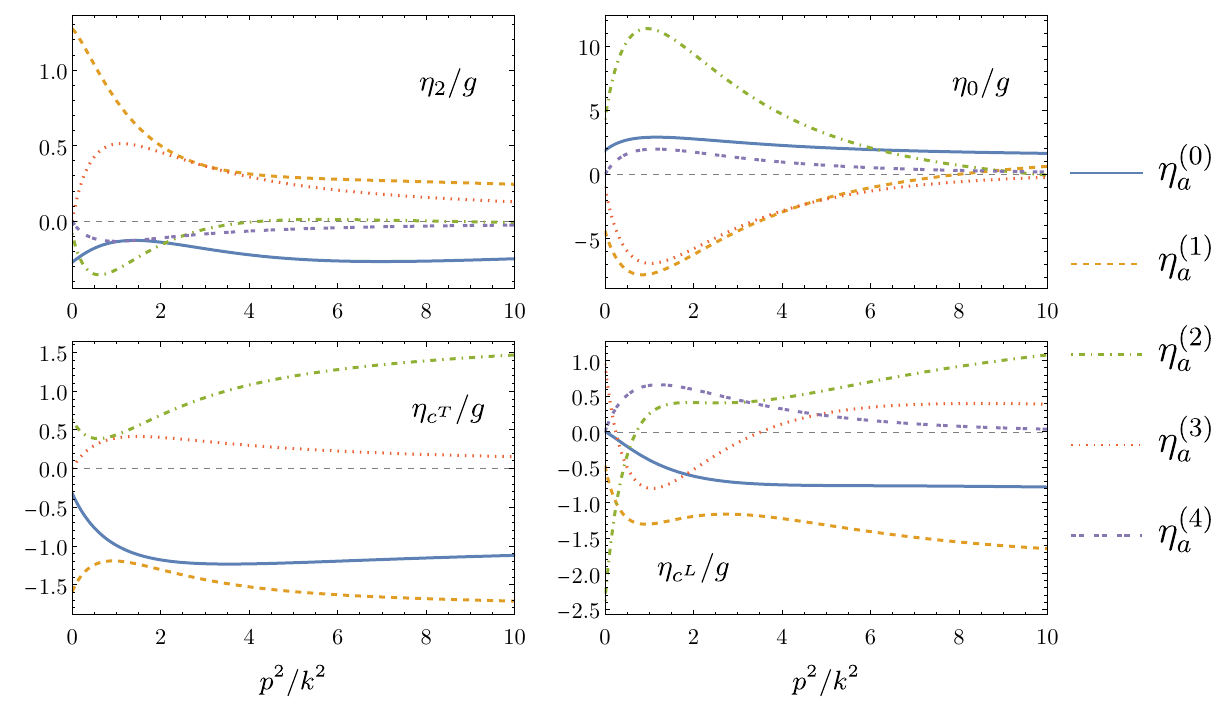}
\caption{\label{fig:eta_gaugedep}Contributions of the different gauge dependence factors $\eta_a^{(i)}$ to the leading-order anomalous dimensions, as defined in \eqref{eq:etagaugedep}.}
\end{figure*}

The coefficients for the spin-two case read
\begin{subequations}\label{eq:eta2}
\begin{align}
    \eta_2^{(0)} = &\frac{\left(\dimlessp^6+27 \dimlessp^4+142 \dimlessp^2+58\right) \dimlessp^4 \Gamma \left(0,\frac{\dimlessp^2}{2}\right)}{240 \pi }-\frac{\left(\dimlessp^6+27 \dimlessp^4+135 \dimlessp^2+29\right) \dimlessp^4 \Gamma \left(0,\dimlessp^2\right)}{120 \pi} \nonumber \\
    &-\frac{255 \dimlessp^6-1093 \dimlessp^4+1424 \dimlessp^2+160}{20 \pi  \dimlessp^8}+\frac{e^{-\dimlessp^2} \left(\dimlessp^{14}+26 \dimlessp^{12}+110 \dimlessp^{10}-58 \dimlessp^8+103 \dimlessp^6+1936 \dimlessp^4+2706 \dimlessp^2-8544\right)}{120 \pi \dimlessp^6} \nonumber \\
    &-\frac{e^{-\frac{\dimlessp^2}{2}} \left(\dimlessp^{16}+25 \dimlessp^{14}+96 \dimlessp^{12}-58 \dimlessp^{10}-372 \dimlessp^8+176 \dimlessp^6+9144 \dimlessp^4-17568 \dimlessp^2-960\right)}{120 \pi  \dimlessp^8} \, , \\
    \eta_2^{(1)} = & \frac{425 \dimlessp^4-2247 \dimlessp^2+2376}{30 \pi  \dimlessp^6}-\frac{\left(4 \dimlessp^6+99 \dimlessp^4+518 \dimlessp^2+554\right) \dimlessp^4 \Gamma\left(0,\frac{\dimlessp^2}{2}\right)}{120 \pi }+\frac{\left(4 \dimlessp^6+99 \dimlessp^4+457 \dimlessp^2+277\right) \dimlessp^4 \Gamma\left(0,\dimlessp^2\right)}{60 \pi } \nonumber \\
    &+\frac{e^{-\frac{\dimlessp^2}{2}} \left(4 \dimlessp^{14}+91 \dimlessp^{12}+352 \dimlessp^{10}+118 \dimlessp^8+304\dimlessp^4+8640 \dimlessp^2-11904\right)}{60 \pi  \dimlessp^6} \nonumber \\
    &-\frac{e^{-\dimlessp^2} \left(4 \dimlessp^{14}+95 \dimlessp^{12}+366 \dimlessp^{10}-6\dimlessp^8+139 \dimlessp^6+1868 \dimlessp^4+2946 \dimlessp^2-7152\right)}{60 \pi  \dimlessp^6} \, , \\
    \eta_2^{(2)} = &\frac{\left(11 \dimlessp^6+252 \dimlessp^4+1372 \dimlessp^2+1596\right) \dimlessp^4 \Gamma\left(0,\frac{\dimlessp^2}{2}\right)}{120 \pi }-\frac{\left(11 \dimlessp^6+252 \dimlessp^4+1142 \dimlessp^2+798\right) \dimlessp^4 \Gamma\left(0,\dimlessp^2\right)}{60 \pi } \nonumber \\
    & -\frac{15 \dimlessp^4-165 \dimlessp^2+164}{5 \pi  \dimlessp^6}-\frac{e^{-\frac{\dimlessp^2}{2}} \left(11 \dimlessp^{14}+230 \dimlessp^{12}+956 \dimlessp^{10}+340 \dimlessp^8+92\dimlessp^6+540 \dimlessp^4+4200 \dimlessp^2-7296\right)}{60 \pi  \dimlessp^6} \nonumber \\
    &+\frac{e^{-\dimlessp^2} \left(11 \dimlessp^{14}+241 \dimlessp^{12}+912\dimlessp^{10}+94 \dimlessp^8+238 \dimlessp^6+912 \dimlessp^4+540 \dimlessp^2-5328\right)}{60 \pi  \dimlessp^6} \, , \\
    \eta_2^{(3)} =& \frac{3 \left(15 \dimlessp^4-21 \dimlessp^2+56\right)}{10 \pi  \dimlessp^6}-\frac{\left(4 \dimlessp^6+87 \dimlessp^4+462 \dimlessp^2+522\right) \dimlessp^4 \Gamma\left(0,\frac{\dimlessp^2}{2}\right)}{40 \pi }+\frac{\left(4 \dimlessp^6+87 \dimlessp^4+373 \dimlessp^2+261\right) \dimlessp^4 \Gamma\left(0,\dimlessp^2\right)}{20 \pi } \nonumber \\
    &+\frac{e^{-\frac{\dimlessp^2}{2}} \left(4 \dimlessp^{14}+79 \dimlessp^{12}+320 \dimlessp^{10}+102 \dimlessp^8+12\dimlessp^6+48 \dimlessp^4-288 \dimlessp^2-1152\right)}{20 \pi  \dimlessp^6} \nonumber \\
    &-\frac{e^{-\dimlessp^2} \left(4 \dimlessp^{14}+83 \dimlessp^{12}+294\dimlessp^{10}+38 \dimlessp^8+59 \dimlessp^6-108 \dimlessp^4-654 \dimlessp^2-816\right)}{20 \pi  \dimlessp^6} \, , \\
    \eta_2^{(4)} = & -\frac{3 \left(5 \dimlessp^2+3\right)}{20 \pi  \dimlessp^4}+\frac{3 \left(\dimlessp^6+21 \dimlessp^4+106 \dimlessp^2+110\right) \dimlessp^4 \Gamma\left(0,\frac{\dimlessp^2}{2}\right)}{80 \pi }-\frac{3 \left(\dimlessp^6+21 \dimlessp^4+85 \dimlessp^2+55\right) \dimlessp^4 \Gamma \left(0,\dimlessp^2\right)}{40\pi } \nonumber \\
    &-\frac{3 e^{-\frac{\dimlessp^2}{2}} \left(\dimlessp^{12}+19 \dimlessp^{10}+72 \dimlessp^8+18 \dimlessp^6+4 \dimlessp^4+16 \dimlessp^2-96\right)}{40 \pi \dimlessp^4} \nonumber \\
    &+\frac{3 e^{-\dimlessp^2} \left(\dimlessp^{12}+20 \dimlessp^{10}+66 \dimlessp^8+6 \dimlessp^6+13 \dimlessp^4-16 \dimlessp^2-90\right)}{40 \pi 
   \dimlessp^4} \, .
\end{align}
\end{subequations}
For the spin-zero case, we have
\begin{subequations}\label{eq:eta0}
\begin{align}
    \eta_0^{(0)} = & -\frac{\left(\dimlessp^6+45 \dimlessp^4+442 \dimlessp^2+838\right) \dimlessp^4 \Gamma \left(0,\frac{\dimlessp^2}{2}\right)}{48 \pi }+\frac{\left(\dimlessp^6+45\dimlessp^4+369 \dimlessp^2+419\right) \dimlessp^4 \Gamma \left(0,\dimlessp^2\right)}{24 \pi } \nonumber \\
    &+\frac{16 \dimlessp^8+31 \dimlessp^6+221 \dimlessp^4-736\dimlessp^2+640}{4 \pi  \dimlessp^8}-\frac{e^{-\dimlessp^2} \left(\dimlessp^{14}+44 \dimlessp^{12}+326 \dimlessp^{10}+134 \dimlessp^8+73 \dimlessp^6-272\dimlessp^4+366 \dimlessp^2+3264\right)}{24 \pi  \dimlessp^6} \nonumber \\
    &+\frac{e^{-\frac{\dimlessp^2}{2}} \left(\dimlessp^{16}+43 \dimlessp^{14}+360 \dimlessp^{12}+266\dimlessp^{10}+54 \dimlessp^8-424 \dimlessp^6-864 \dimlessp^4+5760 \dimlessp^2-3840\right)}{24 \pi  \dimlessp^8} \, , \\
    \eta_0^{(1)} = & \frac{\left(4 \dimlessp^6+153 \dimlessp^4+1274 \dimlessp^2+2078\right) \dimlessp^4 \Gamma \left(0,\frac{\dimlessp^2}{2}\right)}{24 \pi }-\frac{\left(4 \dimlessp^6+153\dimlessp^4+1039 \dimlessp^2+1039\right) \dimlessp^4 \Gamma \left(0,\dimlessp^2\right)}{12 \pi } \nonumber \\
    &+\frac{48 \dimlessp^8-233 \dimlessp^6-1731 \dimlessp^4+4968\dimlessp^2-2880}{6 \pi  \dimlessp^8} \nonumber \\
    &+\frac{e^{-\dimlessp^2} \left(4 \dimlessp^{14}+149 \dimlessp^{12}+894 \dimlessp^{10}+282 \dimlessp^8+217 \dimlessp^6-820\dimlessp^4+966 \dimlessp^2+12144\right)}{12 \pi  \dimlessp^6} \nonumber \\
    &-\frac{e^{-\frac{\dimlessp^2}{2}} \left(4 \dimlessp^{16}+145 \dimlessp^{14}+1000 \dimlessp^{12}+562\dimlessp^{10}+282 \dimlessp^8-860 \dimlessp^6-4320 \dimlessp^4+19200 \dimlessp^2-5760\right)}{12 \pi  \dimlessp^8}\, , \\
    \eta_0^{(2)} = & -\frac{\left(11 \dimlessp^6+360 \dimlessp^4+2692 \dimlessp^2+3948\right) \dimlessp^4 \Gamma \left(0,\frac{\dimlessp^2}{2}\right)}{24 \pi }+\frac{\left(11\dimlessp^6+360 \dimlessp^4+2174 \dimlessp^2+1974\right) \dimlessp^4 \Gamma \left(0,\dimlessp^2\right)}{12 \pi } \nonumber \\
    &-\frac{12 \dimlessp^8-25 \dimlessp^6-1170\dimlessp^4+2696 \dimlessp^2-720}{2 \pi  \dimlessp^8} \nonumber \\
    &-\frac{e^{-\dimlessp^2} \left(11 \dimlessp^{14}+349 \dimlessp^{12}+1836 \dimlessp^{10}+454 \dimlessp^8+478\dimlessp^6-1632 \dimlessp^4+1836 \dimlessp^2+27504\right)}{12 \pi  \dimlessp^6} \nonumber \\
    &+\frac{e^{-\frac{\dimlessp^2}{2}} \left(11 \dimlessp^{16}+338 \dimlessp^{14}+2060\dimlessp^{12}+916 \dimlessp^{10}+557 \dimlessp^8-840 \dimlessp^6-11388 \dimlessp^4+41520 \dimlessp^2-4320\right)}{12 \pi  \dimlessp^8}\, , \\
    \eta_0^{(3)} = & \frac{3 \left(23 \dimlessp^4-337 \dimlessp^2+632\right)}{2 \pi  \dimlessp^6}+\frac{\left(4 \dimlessp^6+117 \dimlessp^4+802 \dimlessp^2+1086\right) \dimlessp^4\Gamma \left(0,\frac{\dimlessp^2}{2}\right)}{8 \pi }-\frac{\left(4 \dimlessp^6+117 \dimlessp^4+643 \dimlessp^2+543\right) \dimlessp^4 \Gamma\left(0,\dimlessp^2\right)}{4 \pi } \nonumber \\
    &-\frac{e^{-\frac{\dimlessp^2}{2}} \left(4 \dimlessp^{14}+109 \dimlessp^{12}+600 \dimlessp^{10}+226 \dimlessp^8+164\dimlessp^6-96 \dimlessp^4-3648 \dimlessp^2+11904\right)}{4 \pi  \dimlessp^6} \nonumber \\
    &+\frac{e^{-\dimlessp^2} \left(4 \dimlessp^{14}+113 \dimlessp^{12}+534\dimlessp^{10}+110 \dimlessp^8+137 \dimlessp^6-444 \dimlessp^4+534 \dimlessp^2+8112\right)}{4 \pi  \dimlessp^6}\, , \\
    \eta_0^{(4)} = & -\frac{3 \left(5 \dimlessp^4-165 \dimlessp^2+336\right)}{4 \pi  \dimlessp^6}-\frac{3 \left(\dimlessp^6+27 \dimlessp^4+174 \dimlessp^2+226\right) \dimlessp^4 \Gamma\left(0,\frac{\dimlessp^2}{2}\right)}{16 \pi }+\frac{3 \left(\dimlessp^6+27 \dimlessp^4+139 \dimlessp^2+113\right) \dimlessp^4 \Gamma \left(0,\dimlessp^2\right)}{8\pi } \nonumber \\
    &+\frac{3 e^{-\frac{\dimlessp^2}{2}} \left(\dimlessp^{14}+25 \dimlessp^{12}+128 \dimlessp^{10}+46 \dimlessp^8+28 \dimlessp^6-16 \dimlessp^4-672\dimlessp^2+2304\right)}{8 \pi  \dimlessp^6} \nonumber \\
    &-\frac{3 e^{-\dimlessp^2} \left(\dimlessp^{14}+26 \dimlessp^{12}+114 \dimlessp^{10}+22 \dimlessp^8+27
   \dimlessp^6-80 \dimlessp^4+138 \dimlessp^2+1632\right)}{8 \pi  \dimlessp^6}\, .
\end{align}
\end{subequations}
For the Faddeev-Popov ghosts, we find for the transverse mode
\begin{subequations}
\begin{align}
    \eta_{c^T}^{(0)} = & \frac{\left(3 \dimlessp^4+70 \dimlessp^2+210\right) \dimlessp^4 \Gamma \left(0,\frac{\dimlessp^2}{2}\right)}{48 \pi }-\frac{\left(3 \dimlessp^4+65\dimlessp^2+105\right) \dimlessp^4 \Gamma \left(0,\dimlessp^2\right)}{24 \pi }-\frac{12 \dimlessp^6+19 \dimlessp^4+47 \dimlessp^2-200}{4 \pi \dimlessp^6} \nonumber \\
    &+\frac{e^{-\dimlessp^2} \left(3 \dimlessp^{12}+62 \dimlessp^{10}+46 \dimlessp^8+7 \dimlessp^6+36 \dimlessp^4+522 \dimlessp^2+720\right)}{24 \pi \dimlessp^6} \nonumber \\
    &-\frac{e^{-\frac{\dimlessp^2}{2}} \left(3 \dimlessp^{12}+64 \dimlessp^{10}+94 \dimlessp^8-76 \dimlessp^6-240 \dimlessp^4+480\dimlessp^2+1920\right)}{24 \pi  \dimlessp^6}\, , \\
    \eta_{c^T}^{(1)} = & -\frac{\left(15 \dimlessp^4+206 \dimlessp^2+442\right) \dimlessp^4 \Gamma \left(0,\frac{\dimlessp^2}{2}\right)}{48 \pi }+\frac{\left(15 \dimlessp^4+181\dimlessp^2+221\right) \dimlessp^4 \Gamma \left(0,\dimlessp^2\right)}{24 \pi }-\frac{72 \dimlessp^6-109 \dimlessp^4+375 \dimlessp^2-72}{12 \pi \dimlessp^6} \nonumber \\
    &-\frac{e^{-\dimlessp^2} \left(15 \dimlessp^{12}+166 \dimlessp^{10}+70 \dimlessp^8+51 \dimlessp^6+4 \dimlessp^4-510 \dimlessp^2-2544\right)}{24 \pi \dimlessp^6} \nonumber \\
    &+\frac{e^{-\frac{\dimlessp^2}{2}} \left(15 \dimlessp^{12}+176 \dimlessp^{10}+150 \dimlessp^8+44 \dimlessp^6+80 \dimlessp^4+1440\dimlessp^2-2688\right)}{24 \pi  \dimlessp^6}\, , \\
    \eta_{c^T}^{(2)} = & \frac{\left(21 \dimlessp^4+250 \dimlessp^2+478\right) \dimlessp^4 \Gamma \left(0,\frac{\dimlessp^2}{2}\right)}{48 \pi }-\frac{\left(21 \dimlessp^4+215\dimlessp^2+239\right) \dimlessp^4 \Gamma \left(0,\dimlessp^2\right)}{24 \pi }+\frac{72 \dimlessp^6-223 \dimlessp^4+621 \dimlessp^2-408}{12 \pi \dimlessp^6} \nonumber \\
    &+\frac{e^{-\dimlessp^2} \left(21 \dimlessp^{12}+194 \dimlessp^{10}+66 \dimlessp^8+65 \dimlessp^6-20 \dimlessp^4-1098 \dimlessp^2-4176\right)}{24\pi  \dimlessp^6} \nonumber \\
    &-\frac{e^{-\frac{\dimlessp^2}{2}} \left(21 \dimlessp^{12}+208 \dimlessp^{10}+146 \dimlessp^8+36 \dimlessp^6+80 \dimlessp^4+1824\dimlessp^2-4992\right)}{24 \pi  \dimlessp^6}\, , \\
    \eta_{c^T}^{(3)} = & -\frac{\left(3 \dimlessp^4+38 \dimlessp^2+82\right) \dimlessp^4 \Gamma \left(0,\frac{\dimlessp^2}{2}\right)}{16 \pi }+\frac{\left(3 \dimlessp^4+33\dimlessp^2+41\right) \dimlessp^4 \Gamma \left(0,\dimlessp^2\right)}{8 \pi }+\frac{25 \dimlessp^4-75 \dimlessp^2+168}{4 \pi  \dimlessp^6} \nonumber \\
    &-\frac{e^{-\dimlessp^2}\left(3 \dimlessp^{12}+30 \dimlessp^{10}+14 \dimlessp^8+7 \dimlessp^6-44 \dimlessp^4-294 \dimlessp^2-816\right)}{8 \pi \dimlessp^6} \nonumber \\
    &+\frac{e^{-\frac{\dimlessp^2}{2}} \left(3 \dimlessp^{12}+32 \dimlessp^{10}+30 \dimlessp^8-4 \dimlessp^6-16 \dimlessp^4+96\dimlessp^2-1152\right)}{8 \pi  \dimlessp^6} \, ,
\end{align}
\end{subequations}
whereas for the longitudinal mode, we have
\begin{subequations}
\begin{align}
    \eta_{c^L}^{(0)} = & -\frac{3 \left(2 \dimlessp^4-6 \dimlessp^2+23\right)}{2 \pi  \dimlessp^4}-\frac{\left(3 \dimlessp^4+28 \dimlessp^2+30\right) \dimlessp^4 \Gamma\left(0,\frac{\dimlessp^2}{2}\right)}{8 \pi }+\frac{\left(3 \dimlessp^4+22 \dimlessp^2+15\right) \dimlessp^4 \Gamma \left(0,\dimlessp^2\right)}{4 \pi} \nonumber \\
    &-\frac{e^{-\dimlessp^2} \left(3 \dimlessp^{10}+19 \dimlessp^8-\dimlessp^6+11 \dimlessp^4+42 \dimlessp^2+54\right)}{4 \pi \dimlessp^4} +\frac{e^{-\frac{\dimlessp^2}{2}} \left(3 \dimlessp^{10}+22 \dimlessp^8-2 \dimlessp^6+8 \dimlessp^4+48 \dimlessp^2+192\right)}{4 \pi  \dimlessp^4} \, , \\
    \eta_{c^L}^{(1)} = & \frac{\left(33 \dimlessp^4+330 \dimlessp^2+454\right) \dimlessp^4 \Gamma \left(0,\frac{\dimlessp^2}{2}\right)}{16 \pi }-\frac{\left(33 \dimlessp^4+263\dimlessp^2+227\right) \dimlessp^4 \Gamma \left(0,\dimlessp^2\right)}{8 \pi }-\frac{24 \dimlessp^6-17 \dimlessp^4-201 \dimlessp^2+600}{4 \pi \dimlessp^6} \nonumber \\
    &+\frac{e^{-\dimlessp^2} \left(33 \dimlessp^{12}+230 \dimlessp^{10}+30 \dimlessp^8+101 \dimlessp^6-44 \dimlessp^4-738 \dimlessp^2-720\right)}{8 \pi \dimlessp^6} \nonumber \\
    &-\frac{e^{-\frac{\dimlessp^2}{2}} \left(33 \dimlessp^{12}+264 \dimlessp^{10}+58 \dimlessp^8+52 \dimlessp^6+320 \dimlessp^4-576\dimlessp^2-1920\right)}{8 \pi  \dimlessp^6} \, , \\
    \eta_{c^L}^{(2)} = & -\frac{\left(57 \dimlessp^4+578 \dimlessp^2+838\right) \dimlessp^4 \Gamma \left(0,\frac{\dimlessp^2}{2}\right)}{16 \pi }+\frac{\left(57 \dimlessp^4+459\dimlessp^2+419\right) \dimlessp^4 \Gamma \left(0,\dimlessp^2\right)}{8 \pi }+\frac{24 \dimlessp^6-125 \dimlessp^4+231 \dimlessp^2-72}{4 \pi \dimlessp^6} \nonumber \\
    &-\frac{e^{-\dimlessp^2} \left(57 \dimlessp^{12}+402 \dimlessp^{10}+74 \dimlessp^8+157 \dimlessp^6-260 \dimlessp^4-642 \dimlessp^2+2544\right)}{8\pi  \dimlessp^6} \nonumber \\
    &+\frac{e^{-\frac{\dimlessp^2}{2}} \left(57 \dimlessp^{12}+464 \dimlessp^{10}+138 \dimlessp^8+116 \dimlessp^6+416 \dimlessp^4-2304\dimlessp^2+2688\right)}{8 \pi  \dimlessp^6} \, , \\
    \eta_{c^L}^{(3)} = & \frac{\left(39 \dimlessp^4+406 \dimlessp^2+618\right) \dimlessp^4 \Gamma \left(0,\frac{\dimlessp^2}{2}\right)}{16 \pi }-\frac{3 \left(13 \dimlessp^4+107\dimlessp^2+103\right) \dimlessp^4 \Gamma \left(0,\dimlessp^2\right)}{8 \pi }+\frac{3 \left(29 \dimlessp^4-139 \dimlessp^2+136\right)}{4 \pi  \dimlessp^6} \nonumber \\
    &+\frac{3e^{-\dimlessp^2} \left(13 \dimlessp^{12}+94 \dimlessp^{10}+22 \dimlessp^8+33 \dimlessp^6-124 \dimlessp^4-90 \dimlessp^2+1392\right)}{8 \pi \dimlessp^6} \nonumber \\
    &-\frac{e^{-\frac{\dimlessp^2}{2}} \left(39 \dimlessp^{12}+328 \dimlessp^{10}+118 \dimlessp^8+140 \dimlessp^6+144 \dimlessp^4-2784\dimlessp^2+4992\right)}{8 \pi  \dimlessp^6} \, , \\
    \eta_{c^L}^{(4)} = & -\frac{3 \left(3 \dimlessp^4+34 \dimlessp^2+58\right) \dimlessp^4 \Gamma \left(0,\frac{\dimlessp^2}{2}\right)}{16 \pi }+\frac{3 \left(3 \dimlessp^4+27\dimlessp^2+29\right) \dimlessp^4 \Gamma \left(0,\dimlessp^2\right)}{8 \pi }-\frac{3 \left(5 \dimlessp^4-81 \dimlessp^2+168\right)}{4 \pi  \dimlessp^6} \nonumber \\
    &-\frac{3e^{-\dimlessp^2} \left(3 \dimlessp^{12}+24 \dimlessp^{10}+8 \dimlessp^8+7 \dimlessp^6-32 \dimlessp^4+114 \dimlessp^2+816\right)}{8 \pi  \dimlessp^6} \nonumber \\
    &+\frac{3e^{-\frac{\dimlessp^2}{2}} \left(3 \dimlessp^{12}+28 \dimlessp^{10}+14 \dimlessp^8+12 \dimlessp^6-16 \dimlessp^4-288 \dimlessp^2+1152\right)}{8 \pi \dimlessp^6} \, .
\end{align}
\end{subequations}
In these expressions, $\Gamma(x,y)$ denotes the incomplete Gamma function.

\section{Truncation check}

To test whether neglecting the anomalous dimensions in loop diagrams is a good approximation, we compare the full, numerical solutions to our approximation in \autoref{fig:eta_comparison} for $\beta=-1$, for different values of $g$.\footnote{In preparing this work, we noticed that the numerical results in~\cite{Knorr:2021niv} were off by a factor of $1/2$.} As expected, for small values of $g$, the difference is completely negligible, since the anomalous dimensions in loops are higher-loop effects, controlled by $g$. Surprisingly, even for rather large values of $g$, the gravitational anomalous dimensions $\eta_{2,0}$ are well approximated by the leading term. In the ghost sector, noticeable quantitative deviations start at around $g=1$, but remain within about 30\% for $g=2$. For our choice of $g_\ast$ in the analytical results above, we are well within the accurate regime of our approximation.

\begin{figure*}[t]
\centering
\includegraphics[width=0.8\linewidth]{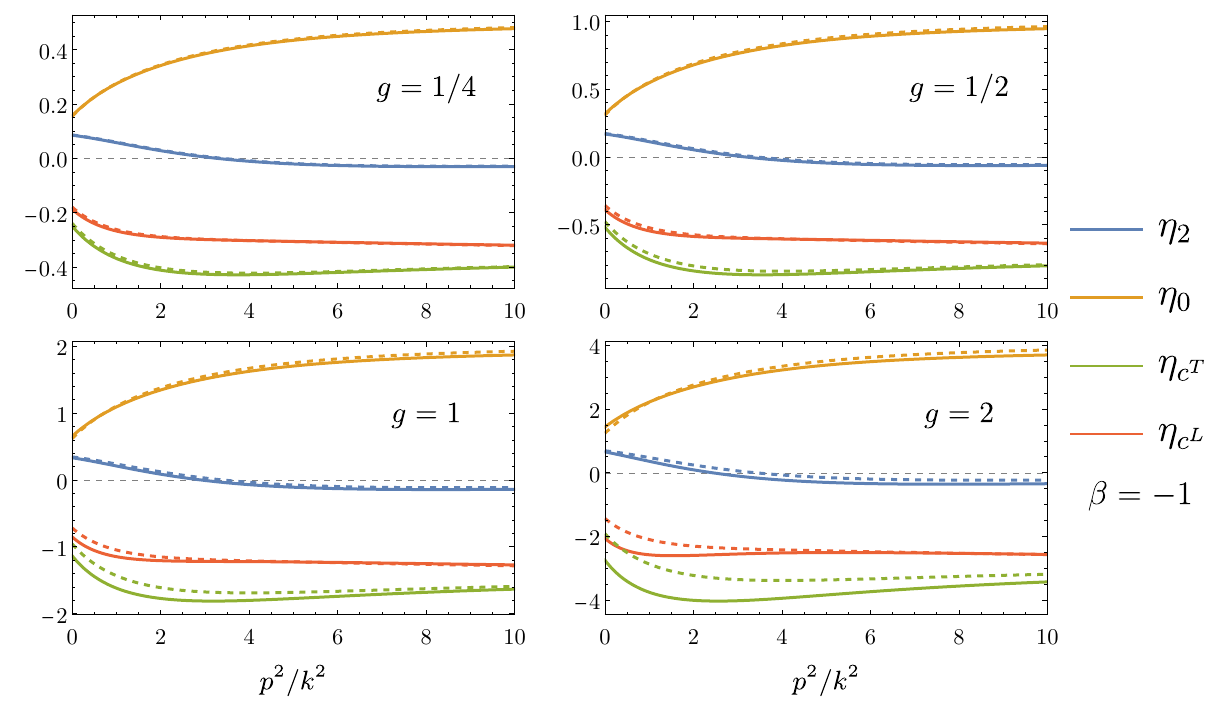}
\caption{\label{fig:eta_comparison}Comparison of the full numerical solution for the anomalous dimensions (solid lines) with the leading-order solutions (dashed lines) for $\beta=-1$ and different values of $g$. The graviton anomalous dimensions agree even at large values of $g$, whereas the ghost anomalous dimensions start to deviate quantitatively at around $g=1$.}
\end{figure*}

More generally, the approximate anomalous dimensions above agree with the numerical solutions of the full equations both qualitatively and quantitatively, for small values of $g$. The value of $g$ up until which the approximation is reliable depends on the chosen gauge. For the popular choice $\beta=1$, the spin-zero anomalous dimension starts to deviate significantly from the leading-order solution already around $g=1$. This is shown in \autoref{fig:eta_comparison_betaone}. Consequently, in this gauge choice, the feedback has to be taken into account self-consistently in that regime of coupling values.

\begin{figure*}[t]
\centering
\includegraphics[width=0.8\linewidth]{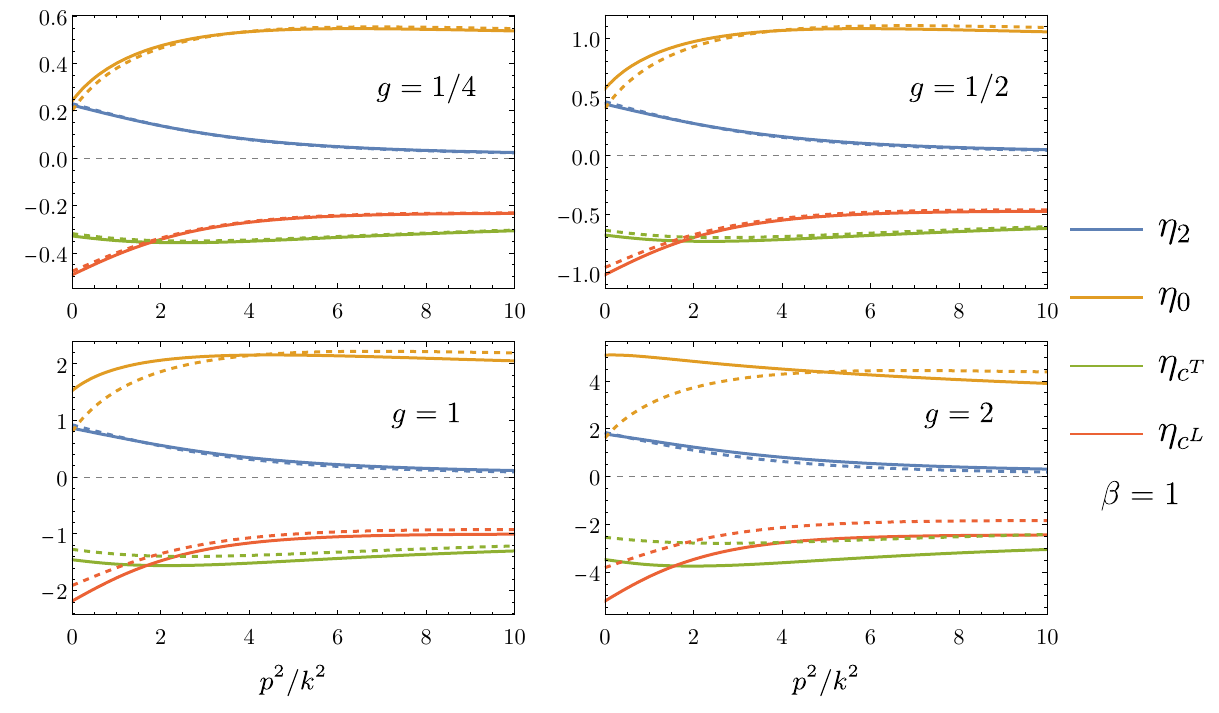}
\caption{\label{fig:eta_comparison_betaone}Comparison of the full numerical solution for the anomalous dimensions (solid lines) with the leading-order solutions (dashed lines) for $\beta=1$ and different values of $g$. The spin-two graviton anomalous dimension agrees even at large values of $g$, whereas all other modes start to deviate quantitatively (or even qualitatively for the spin-zero mode) at around $g=1$.}
\end{figure*}

\section{Wave function renormalisations}

To compute the propagators from the anomalous dimensions, we have to integrate the latter. By definition, the anomalous dimensions are related to the wave function renormalisations $Z$ via
\begin{equation}\label{eq:defeta}
    k \partial_k Z_k(p^2) = - \eta_k(p^2) \, .
\end{equation}
The wave function renormalisations in turn are the dressing of the propagators,
\begin{equation}
    G(p^2) = \frac{1}{p^2 Z_{k=0}(p^2)} \, .
\end{equation}
Thus, to compute the propagator, we have to integrate \eqref{eq:defeta},
\begin{equation}
    Z_k(p^2) = Z_\Lambda(p^2) e^{\int_k^\Lambda \text{d}k' \, \frac{\eta_{k'}(p^2)}{k'}} \, ,
\end{equation}
and take the limit $k\to0$ as well as $\Lambda\to\infty$. To arrive at a finite result for $Z_{k=0}$, we have to impose a renormalisation condition on $Z_\Lambda$~\cite{Bonanno:2021squ}. This is necessary since the integral in the exponent diverges in the limit $\Lambda\to\infty$. We follow the ideas of~\cite{Knorr:2021niv} and choose
\begin{equation}\label{eq:solZ}
    Z_{k=0}(p^2) = e^{\int_0^\infty \text{d}k' \, \frac{\eta_{k'}(p^2)-\eta_{k'}(0)}{k'}} \, .
\end{equation}
This choice removes only the divergent term while also imposing $Z_{k=0}(0)=1$.

Due to the structure of \eqref{eq:solZ}, we report the results for $\ln Z_a$ instead of for $Z_a$, where the subscript indicates the mode. It is convenient to parameterise the results as
\begin{equation}
    \ln Z_a(p^2) = g_\ast \, \sum_{n=1}^7 c_a^{(n)}(\physdimlesspsq) f^{(n)}(\physdimlesspsq)  \, ,
\end{equation}
where the $f^{(n)}$ are gauge-independent functions, the $c_a^{(n)}$ carry the gauge and mode dependence, and all momentum dependence comes in terms of the dimensionless combination
\begin{equation}
    \physdimlesspsq = \frac{G_N p^2}{g_\ast} \, .
\end{equation}
The functions $f^{(n)}$ are given by
\begin{subequations}
\begin{align}
    f^{(1)}(X) = &X \left(4 \, _3F_3(1,1,1;2,2,2;X)-\, _3F_3\left(1,1,1;2,2,2;\frac{X}{2}\right)\right)+\frac{1}{4} \left(-8 \gamma  \text{Ei}(X)-\pi ^2+2 \gamma^2\right)-\frac{21}{X^4}-\frac{4}{X^3}-\frac{1}{X^2} \nonumber \\
    &+e^{X/2} \left(\frac{96}{X^4}+\frac{16}{X^3}+\frac{4}{X^2}+\frac{2}{X}\right) \Gamma \left(0,\frac{X}{2}\right)+e^X\left(-\frac{48}{X^5}-\frac{12}{X^4}-\frac{4}{X^3}-\frac{2}{X^2}-\frac{2}{X}\right) \Gamma (0,X) \nonumber \\
    &-\frac{1}{2} \left[\ln \left(\frac{X}{2}\right)\right]^2 - \left[\ln(X)\right]^2+(\ln (X)-\gamma ) \ln\left(\frac{X}{2}\right)+3 \gamma  \ln (X)+\ln (-X) \left(-\ln \left(\frac{X}{2}\right)+2 \ln (X)-\gamma \right) \nonumber \\
    &+\left(-\ln \left(\frac{X}{2}\right)-\gamma \right) \Gamma\left(0,-\frac{X}{2}\right)+2 \ln (X) \Gamma (0,-X) + {}_1F_1^{(2,0,0)}\left(0;1\big|X\right) - \frac{1}{2} {}_1F_1^{(2,0,0)}\left(0;1\Bigg|\frac{X}{2}\right) \, , \\
    f^{(2)}(X) = &2 X \, _3F_3\left(1,1,1;2,2,2\bigg|\frac{X}{2}\right)+\frac{4}{X^2}+e^{X/2} \left(-\frac{8}{X^2}-\frac{4}{X}\right) \Gamma \left(0,\frac{X}{2}\right)+\frac{4}{X}+\left[\ln\left(\frac{X}{2}\right)\right]^2 \nonumber \\
    & +(2 \gamma -2 \ln (X)) \ln \left(\frac{X}{2}\right)+\ln (-X) \left(2 \ln \left(\frac{X}{2}\right)+2 \gamma \right)-2 \gamma  \ln (X) \nonumber \\
    & +\left(2\ln \left(\frac{X}{2}\right)+2 \gamma \right) \Gamma \left(0,-\frac{X}{2}\right)-\frac{\pi ^2}{2}+\gamma ^2 + {}_1F_1^{(2,0,0)}\left(0;1\Bigg|\frac{X}{2}\right) \, , \\
    f^{(3)}(X) =& \gamma + \ln X \, , \\
    f^{(4)}(X) =& \gamma + \ln \frac{X}{2} \, , \\
    f^{(5)}(X) =& e^X \Gamma(0,X) \, , \\
    f^{(6)}(X) =& e^\frac{X}{2} \Gamma\left(0,\frac{X}{2}\right) \, , \\
    f^{(7)}(X) =& 1 \, .
\end{align}
\end{subequations}
In this, $\gamma$ is the Euler-Mascheroni constant, and $\text{Ei}$ is the exponential integral function. Moreover, the derivative of the hypergeometric function with respect to its first argument is to be understood as
\begin{equation}
    {}_1F_1^{(2,0,0)}\left(0;1|z\right) = \lim_{a\to0} \frac{\partial^2}{\partial a^2} {}_1F_1\left(a;1|z\right) \, .
\end{equation}
Note that, as written above, all branch cuts combine in a way that only a single branch cut on the negative real axis of $X$ remains.

The coefficients $c_a^{(n)}$ are given as
\begin{subequations}
\begin{align}
    c_2^{(1)} =& \frac{X^5 \left(3 \tilde{\beta }^2-4 \tilde{\beta }+1\right)^2}{480 \pi }-\frac{3 X^4 \left(\tilde{\beta }-1\right)^2 \left(3 \tilde{\beta }-1\right) \left(7 \tilde{\beta}-3\right)}{160 \pi } \nonumber \\
    &+\frac{X^3 \left(\tilde{\beta } \left(\tilde{\beta } \left(3 \tilde{\beta } \left(255 \tilde{\beta }-746\right)+2284\right)-914\right)+135\right)}{480 \pi}-\frac{X^2 \left(\tilde{\beta }-1\right) \left(3 \tilde{\beta } \left(3 \tilde{\beta } \left(55 \tilde{\beta }-119\right)+175\right)-29\right)}{480 \pi } \, , \\
    c_2^{(2)} =& \frac{X^2 \left(\tilde{\beta }-1\right) \left(3 \tilde{\beta } \left(3 \tilde{\beta } \left(55 \tilde{\beta }-119\right)+175\right)-29\right)}{960 \pi }-\frac{X^3\left(\tilde{\beta }-1\right) \left(3 \tilde{\beta }-1\right) \left(\tilde{\beta } \left(63 \tilde{\beta }-94\right)+7\right)}{960 \pi } \, , \\
    c_2^{(3)} =& \frac{\tilde{\beta } \left(\tilde{\beta } \left(18 \left(\tilde{\beta }-1\right) \tilde{\beta }-121\right)+234\right)-118}{30 \pi }+\frac{2 \left(\tilde{\beta }-1\right) \left(51\tilde{\beta }^2-60 \tilde{\beta }+89\right)}{5 \pi  X^3} \nonumber \\
    & +\frac{3 \left(\tilde{\beta } \left(\tilde{\beta } \left(9 \tilde{\beta } \left(5 \tilde{\beta}+6\right)-652\right)+1122\right)-625\right)}{40 \pi  X^2}+\frac{\tilde{\beta } \left(3 \tilde{\beta } \left(3 \tilde{\beta } \left(37 \tilde{\beta}-46\right)-764\right)+4654\right)-2521}{120 \pi  X} \, , \\
    c_2^{(4)} =& \frac{4 \tilde{\beta } \left(\tilde{\beta } \left(499-72 \left(\tilde{\beta }-1\right) \tilde{\beta }\right)-1176\right)+2093}{480 \pi }+\frac{4 \left(89-4 \tilde{\beta }\left(\tilde{\beta } \left(9 \tilde{\beta }-19\right)+31\right)\right)}{5 \pi  X^3} \nonumber \\
    & +\frac{371-\tilde{\beta } \left(3 \tilde{\beta } \left(6 \tilde{\beta } \left(\tilde{\beta}+2\right)-109\right)+608\right)}{5 \pi  X^2}+\frac{4 \tilde{\beta } \left(3 \tilde{\beta } \left(6 \left(1-2 \tilde{\beta }\right) \tilde{\beta}+103\right)-688\right)+1643}{60 \pi  X}-\frac{4}{\pi  X^4} \, , \\
    c_2^{(5)} =& -\frac{\left(\tilde{\beta }-1\right) \left(3 \tilde{\beta } \left(\tilde{\beta } \left(33 \tilde{\beta }-85\right)+51\right)-77\right)}{2 \pi  X^3}+\frac{\tilde{\beta }\left(\tilde{\beta } \left(3 \tilde{\beta } \left(45 \tilde{\beta }-134\right)+368\right)-106\right)+3}{2 \pi  X^2} \nonumber \\
    & +\frac{\tilde{\beta } \left(\tilde{\beta } \left(3 \left(8-3\tilde{\beta }\right) \tilde{\beta }+2\right)-28\right)+17}{2 \pi  X} \, , \\
    c_2^{(6)} =& \frac{\tilde{\beta } \left(\tilde{\beta } \left(3 \tilde{\beta } \left(207 \tilde{\beta }-562\right)+1561\right)-551\right)-5}{30 \pi }+\frac{366-16 \tilde{\beta }\left(\tilde{\beta } \left(9 \tilde{\beta }-19\right)+31\right)}{5 \pi  X^3} \nonumber \\
    &+\frac{2 \tilde{\beta } \left(\tilde{\beta } \left(9 \tilde{\beta } \left(53 \tilde{\beta}-170\right)+1771\right)-914\right)+439}{10 \pi  X^2}+\frac{\tilde{\beta } \left(5006-3 \tilde{\beta } \left(3 \tilde{\beta } \left(457 \tilde{\beta}-1322\right)+4081\right)\right)-803}{30 \pi  X}-\frac{4}{\pi  X^4} \, , \\
    c_2^{(7)} =& \frac{8517-4 \tilde{\beta } \left(\tilde{\beta } \left(3 \tilde{\beta } \left(2853 \tilde{\beta }-7898\right)+14810\right)+758\right)}{5760 \pi }-\frac{\tilde{\beta }\left(\tilde{\beta } \left(3 \tilde{\beta } \left(1155 \tilde{\beta }-3334\right)+8932\right)-2278\right)+243}{160 \pi  X^2} \nonumber \\
    &+\frac{3 \tilde{\beta } \left(3 \tilde{\beta }\left(3 \tilde{\beta } \left(1481 \tilde{\beta }-4374\right)+15284\right)-24010\right)+15865}{1440 \pi  X}+\frac{2}{\pi  X^3} \, ,
\end{align}
\end{subequations}
for the spin-two sector,
\begin{subequations}
\begin{align}
    c_0^{(1)} =& -\frac{X^5 \left(3 \tilde{\beta }^2-4 \tilde{\beta }+1\right)^2}{96 \pi }+\frac{3 X^4 \left(\tilde{\beta }-1\right)^2 \left(3 \tilde{\beta }-1\right) \left(9 \tilde{\beta}-5\right)}{32 \pi } \nonumber \\
    &+\frac{X^3 \left(\tilde{\beta } \left(\tilde{\beta } \left(3 \left(1286-417 \tilde{\beta }\right) \tilde{\beta }-4348\right)+2078\right)-369\right)}{96 \pi} \nonumber \\
    &+\frac{X^2 \left(\tilde{\beta } \left(3 \tilde{\beta } \left(3 \tilde{\beta } \left(113 \tilde{\beta }-362\right)+1316\right)-2078\right)+419\right)}{96 \pi } \, , \\
    c_0^{(2)} =& \frac{X^3 \left(\tilde{\beta }-1\right)^2 \left(3 \tilde{\beta }-1\right) \left(105 \tilde{\beta }-73\right)}{192 \pi }+\frac{X^2 \left(\tilde{\beta } \left(2078-3 \tilde{\beta }\left(3 \tilde{\beta } \left(113 \tilde{\beta }-362\right)+1316\right)\right)-419\right)}{192 \pi } \, , \\
    c_0^{(3)} =& \frac{\tilde{\beta } \left(\tilde{\beta } \left(54 \left(7-2 \tilde{\beta }\right) \tilde{\beta }-389\right)+126\right)-2}{6 \pi }-\frac{2 \left(\tilde{\beta }-1\right) \left(3\tilde{\beta }-2\right) \left(51 \tilde{\beta }^2-84 \tilde{\beta }+17\right)}{\pi  X^3} \nonumber \\
    &+\frac{3 \tilde{\beta } \left(\tilde{\beta } \left(3 \left(842-249 \tilde{\beta}\right) \tilde{\beta }-2852\right)+1242\right)-483}{8 \pi  X^2}+\frac{\tilde{\beta } \left(4286-3 \tilde{\beta } \left(897 \tilde{\beta }^2-3078 \tilde{\beta}+3428\right)\right)-497}{24 \pi  X} \, , \\
    c_0^{(4)} =& \frac{\tilde{\beta } \left(\tilde{\beta } \left(144 \tilde{\beta } \left(12 \tilde{\beta }-41\right)+5291\right)-972\right)-56}{96 \pi }+\frac{20 \left(2-3 \tilde{\beta}\right)^2}{\pi  X^4}+\frac{4 \left(3 \tilde{\beta }-2\right) \left(\tilde{\beta } \left(4 \tilde{\beta } \left(9 \tilde{\beta }-25\right)+85\right)-20\right)}{\pi X^3} \nonumber \\
    &+\frac{2 \left(\tilde{\beta } \left(3 \tilde{\beta } \left(\tilde{\beta } \left(57 \tilde{\beta }-200\right)+227\right)-305\right)+44\right)}{\pi  X^2}+\frac{\tilde{\beta} \left(9 \tilde{\beta } \left(8 \tilde{\beta } \left(19 \tilde{\beta }-67\right)+563\right)-1910\right)+202}{12 \pi  X} \, , \\
    c_0^{(5)} =& \frac{\tilde{\beta } \left(3 \tilde{\beta } \left(5 \tilde{\beta } \left(27 \tilde{\beta }-82\right)+552\right)-1066\right)+283}{2 \pi  X^3}+\frac{\tilde{\beta }\left(\tilde{\beta } \left(15 \left(194-63 \tilde{\beta }\right) \tilde{\beta }-3208\right)+1478\right)-249}{2 \pi  X^2} \nonumber \\
    &+\frac{\tilde{\beta } \left(\tilde{\beta } \left(15\tilde{\beta } \left(3 \tilde{\beta }-8\right)+98\right)-28\right)-1}{2 \pi  X} \, , \\
    c_0^{(6)} =& \frac{\tilde{\beta } \left(\tilde{\beta } \left(36 \left(234-83 \tilde{\beta }\right) \tilde{\beta }-8341\right)+3470\right)-469}{24 \pi }+\frac{20 \left(2-3 \tilde{\beta}\right)^2}{\pi  X^4}+\frac{2 \left(3 \tilde{\beta }-2\right) \left(\tilde{\beta } \left(8 \tilde{\beta } \left(9 \tilde{\beta }-25\right)+155\right)-30\right)}{\pi X^3} \nonumber \\
    &-\frac{\tilde{\beta } \left(\tilde{\beta } \left(6 \tilde{\beta } \left(297 \tilde{\beta }-934\right)+6947\right)-3796\right)+802}{2 \pi  X^2}+\frac{\tilde{\beta } \left(3\tilde{\beta } \left(3 \tilde{\beta } \left(719 \tilde{\beta }-2202\right)+7310\right)-10175\right)+1742}{6 \pi  X} \, , \\
    c_0^{(7)} =& \frac{\tilde{\beta } \left(\tilde{\beta } \left(36 \tilde{\beta } \left(1705 \tilde{\beta }-5202\right)+214247\right)-101212\right)+14952}{1152 \pi }-\frac{10 \left(2-3\tilde{\beta }\right)^2}{\pi  X^3} \nonumber \\
    &+\frac{\tilde{\beta } \left(\tilde{\beta } \left(9 \tilde{\beta } \left(1111 \tilde{\beta }-3494\right)+35548\right)-16498\right)+2709}{32\pi  X^2} \nonumber \\
    &-\frac{3 \tilde{\beta } \left(3 \tilde{\beta } \left(3 \tilde{\beta } \left(1831 \tilde{\beta }-5562\right)+18308\right)-26678\right)+15679}{288 \pi  X} \, ,
\end{align}
\end{subequations}
for the spin-zero sector,
\begin{subequations}
\begin{align}
    c_{c^T}^{(1)} =& \frac{X^4 \left(\tilde{\beta }-1\right)^2 \left(3 \tilde{\beta }-1\right)}{32 \pi }-\frac{X^3 \left(\tilde{\beta }-1\right) \left(\tilde{\beta } \left(99 \tilde{\beta}-116\right)+65\right)}{96 \pi }+\frac{X^2 \left(\tilde{\beta }-1\right) \left(\tilde{\beta } \left(123 \tilde{\beta }-116\right)+105\right)}{96 \pi } \, , \\
    c_{c^T}^{(2)} =& \frac{5 X^3 \left(\tilde{\beta }-1\right)^2 \left(3 \tilde{\beta }-1\right)}{192 \pi }-\frac{X^2 \left(\tilde{\beta }-1\right) \left(\tilde{\beta } \left(123 \tilde{\beta}-116\right)+105\right)}{192 \pi } \, , \\
    c_{c^T}^{(3)} =& \frac{\tilde{\beta } \left(\tilde{\beta } \left(15 \tilde{\beta }-29\right)+27\right)-14}{6 \pi }+\frac{\tilde{\beta } \left(51 \tilde{\beta }^2-87 \tilde{\beta}+53\right)+15}{\pi  X^3}+\frac{3 \tilde{\beta } \left(3 \tilde{\beta } \left(29 \tilde{\beta }-57\right)+113\right)+33}{8 \pi  X^2} \nonumber \\
    &+\frac{\tilde{\beta } \left(\tilde{\beta }\left(237 \tilde{\beta }-505\right)+379\right)-63}{24 \pi  X} \, , \\
    c_{c^T}^{(4)} =& \frac{\tilde{\beta } \left(\left(41-15 \tilde{\beta }\right) \tilde{\beta }-30\right)+8}{6 \pi }-\frac{8 \left(\tilde{\beta } \left(\tilde{\beta } \left(9 \tilde{\beta}-13\right)+7\right)+5\right)}{\pi  X^3}-\frac{2 \left(\tilde{\beta } \left(3 \tilde{\beta } \left(7 \tilde{\beta }-15\right)+29\right)+5\right)}{\pi  X^2} \nonumber \\
    &+\frac{\tilde{\beta} \left(13 \left(7-3 \tilde{\beta }\right) \tilde{\beta }-61\right)+15}{3 \pi  X} \, , \\
    c_{c^T}^{(5)} =& \frac{\tilde{\beta } \left(\tilde{\beta } \left(225 \tilde{\beta }-413\right)+327\right)-75}{2 \pi  X^3}+\frac{17-3 \tilde{\beta } \left(\tilde{\beta } \left(35 \tilde{\beta}-67\right)+49\right)}{2 \pi  X^2}+\frac{6 \left(\tilde{\beta }-1\right) \tilde{\beta }+3}{\pi  X} \, , \\
    c_{c^T}^{(6)} =& -\frac{5 \tilde{\beta } \left(\tilde{\beta } \left(3 \tilde{\beta }-7\right)+5\right)+4}{6 \pi }-\frac{8 \left(\tilde{\beta } \left(\tilde{\beta } \left(9 \tilde{\beta}-13\right)+7\right)+5\right)}{\pi  X^3}+\frac{\tilde{\beta } \left(\left(277-129 \tilde{\beta }\right) \tilde{\beta }-251\right)+115}{\pi  X^2} \nonumber \\
    &+\frac{\tilde{\beta }\left(\tilde{\beta } \left(465 \tilde{\beta }-1061\right)+875\right)-255}{6 \pi  X} \, , \\
    c_{c^T}^{(7)} =& \frac{\tilde{\beta } \left(145 \tilde{\beta } \left(3 \tilde{\beta }-7\right)+413\right)-361}{288 \pi }+\frac{\left(\tilde{\beta }-1\right) \left(\tilde{\beta } \left(381\tilde{\beta }-172\right)+575\right)}{32 \pi  X^2} \nonumber \\
    &+\frac{\tilde{\beta } \left(\left(3931-1911 \tilde{\beta }\right) \tilde{\beta }-3265\right)+1149}{96 \pi  X} \, ,
\end{align}
\end{subequations}
for the transverse ghost, and
\begin{subequations}
\begin{align}
    c_{c^L}^{(1)} =& \frac{3 X^4 \left(\tilde{\beta }-2\right) \left(\tilde{\beta }-1\right)^2 \left(3 \tilde{\beta }-1\right)}{32 \pi }-\frac{X^3 \left(\tilde{\beta }-1\right) \left(3 \tilde{\beta }\left(\tilde{\beta } \left(27 \tilde{\beta }-80\right)+73\right)-44\right)}{32 \pi } \nonumber \\
    &+\frac{X^2 \left(\tilde{\beta }-1\right) \left(\tilde{\beta } \left(3 \tilde{\beta }\left(29 \tilde{\beta }-74\right)+197\right)-30\right)}{32 \pi } \, , \\
    c_{c^L}^{(2)} =& \frac{X^3 \left(\tilde{\beta }-1\right)^2 \left(3 \tilde{\beta }-1\right) \left(7 \tilde{\beta }-12\right)}{64 \pi }-\frac{X^2 \left(\tilde{\beta }-1\right) \left(\tilde{\beta }\left(3 \tilde{\beta } \left(29 \tilde{\beta }-74\right)+197\right)-30\right)}{64 \pi } \, , \\
    c_{c^L}^{(3)} =& -\frac{\tilde{\beta } \left(\tilde{\beta } \left(15 \left(\tilde{\beta }-3\right) \tilde{\beta }+41\right)-13\right)+1}{2 \pi }-\frac{3 \tilde{\beta } \left(\tilde{\beta }\left(51 \tilde{\beta }^2-87 \tilde{\beta }+53\right)+15\right)}{\pi  X^3} \nonumber \\
    &+\frac{9 \tilde{\beta } \left(\tilde{\beta } \left(13 \left(19-9 \tilde{\beta }\right) \tilde{\beta}-177\right)+1\right)+54}{8 \pi  X^2}+\frac{\tilde{\beta } \left(\tilde{\beta } \left(-393 \tilde{\beta }^2+993 \tilde{\beta }-827\right)+167\right)+12}{8 \pi  X} \, , \\
    c_{c^L}^{(4)} =& \frac{\tilde{\beta } \left(\tilde{\beta } \left(6 \tilde{\beta } \left(5 \tilde{\beta }-16\right)+109\right)-35\right)-4}{4 \pi }+\frac{24 \tilde{\beta } \left(\tilde{\beta }\left(\tilde{\beta } \left(9 \tilde{\beta }-13\right)+7\right)+5\right)}{\pi  X^3} \nonumber \\
    &+\frac{6 \left(\tilde{\beta } \left(\tilde{\beta } \left(\tilde{\beta } \left(27 \tilde{\beta}-55\right)+38\right)+4\right)-4\right)}{\pi  X^2}+\frac{\tilde{\beta } \left(\tilde{\beta } \left(3 \tilde{\beta } \left(17 \tilde{\beta}-45\right)+119\right)-23\right)-6}{\pi  X} \, , \\
    c_{c^L}^{(5)} =& \frac{90-3 \tilde{\beta } \left(\tilde{\beta } \left(15 \tilde{\beta } \left(\tilde{\beta }+9\right)-313\right)+257\right)}{2 \pi  X^3}-\frac{9 \tilde{\beta } \left(\tilde{\beta} \left(5 \tilde{\beta } \left(3 \tilde{\beta }-17\right)+127\right)-79\right)+96}{2 \pi  X^2}-\frac{9 \left(2 \left(\tilde{\beta }-1\right) \tilde{\beta }+1\right)}{\pi  X} \, , \\
    c_{c^L}^{(6)} =& \frac{\tilde{\beta } \left(\tilde{\beta } \left(\left(85-21 \tilde{\beta }\right) \tilde{\beta }-107\right)+55\right)-9}{2 \pi }+\frac{24 \tilde{\beta } \left(\tilde{\beta }\left(\tilde{\beta } \left(9 \tilde{\beta }-13\right)+7\right)+5\right)}{\pi  X^3} \nonumber \\
    &-\frac{3 \left(\tilde{\beta } \left(\tilde{\beta } \left(\tilde{\beta } \left(69 \tilde{\beta}-251\right)+371\right)-215\right)+38\right)}{\pi  X^2}+\frac{\left(\tilde{\beta }-1\right) \left(\tilde{\beta } \left(3 \tilde{\beta } \left(131 \tilde{\beta}-414\right)+1145\right)-246\right)}{2 \pi  X} \, , \\
    c_{c^L}^{(7)} =& \frac{\tilde{\beta } \left(\tilde{\beta } \left(\tilde{\beta } \left(1131 \tilde{\beta }-3317\right)+3391\right)-1535\right)+282}{96 \pi }+\frac{3 \left(\tilde{\beta }-1\right)\left(\tilde{\beta } \left(\tilde{\beta } \left(1089 \tilde{\beta }-2194\right)+1539\right)-210\right)}{32 \pi  X^2} \nonumber \\
    &-\frac{\tilde{\beta } \left(\tilde{\beta } \left(525\tilde{\beta }^2-3357 \tilde{\beta }+6175\right)-4075\right)+636}{32 \pi  X} \, ,
\end{align}
\end{subequations}
for the longitudinal ghost.

All these results are also available in a supplemental Mathematica notebook.

\section{Expansion of form factors for small momenta}

In this appendix, we provide the small-momentum expansion of the form factors $F_{C,R}(p^2)$. This expansion can be used to read off Wilson coefficients. We find
\begin{equation}\label{eq:wc}
\begin{aligned}
    \frac{F_{C,R}(p^2)}{G_N} \sim & \left[ q_1^{C,R} + q_2^{C,R} \ln \frac{X}{2} \right] \\
    &+ X \, \left[ \frac{g_\ast}{2} \left(q_1^{C,R} + q_2^{C,R} \ln \frac{X}{2}\right)^2 + q_3^{C,R} \left( \left(\ln X\right)^2 - \left( \ln \frac{X}{2} \right)^2 \right) + q_4^{C,R} \ln X + q_5^{C,R} \ln \frac{X}{2} + q_6^{C,R} \right] + \dots \, .
\end{aligned}
\end{equation}
Here, we again used $X=G_N p^2/g_\ast$, and introduced the coefficients
\begin{subequations}
\begin{align}
    q_1^C &= \frac{22500 \tilde{\beta }^4-71100 \tilde{\beta }^3+83700 \tilde{\beta }^2-53300 \tilde{\beta }+240 \gamma  \left(90 \tilde{\beta }^4-315 \tilde{\beta }^3+255 \tilde{\beta }^2+55 \tilde{\beta }-43\right)+419}{28800 \pi } \, , \\
    q_2^C &= \frac{90 \tilde{\beta }^4-315 \tilde{\beta }^3+255 \tilde{\beta }^2+55 \tilde{\beta }-43}{120 \pi } \, , \\
    q_3^C &= \frac{495 \tilde{\beta }^4-1566 \tilde{\beta }^3+1596 \tilde{\beta }^2-554 \tilde{\beta }+29}{480 \pi } \, , \\
    q_4^C &= \frac{35325 \tilde{\beta }^4-101898 \tilde{\beta }^3+102228 \tilde{\beta }^2-37502 \tilde{\beta }+60 \gamma  \left(495 \tilde{\beta }^4-1566 \tilde{\beta }^3+1596 \tilde{\beta}^2-554 \tilde{\beta }+29\right)+4847}{14400 \pi } \, , \\
    q_5^C &= -\frac{67950 \tilde{\beta }^4-193716 \tilde{\beta }^3+191781 \tilde{\beta }^2-74944 \tilde{\beta }+120 \gamma  \left(495 \tilde{\beta }^4-1566 \tilde{\beta }^3+1596 \tilde{\beta}^2-554 \tilde{\beta }+29\right)+12384}{28800 \pi } \, , \\
    q_6^C &= \frac{-98100 \tilde{\beta }^4+296544 \tilde{\beta }^3-217285 \tilde{\beta }^2-31252 \tilde{\beta }+20 \gamma  \left(540 \tilde{\beta }^4-2016 \tilde{\beta }^3+2535 \tilde{\beta}^2-12 \tilde{\beta }-538\right)+36417}{115200 \pi } \, ,
\end{align}
\end{subequations}
for $F_C$, and
\begin{subequations}
\begin{align}
    q_1^R &= -\frac{46800 \tilde{\beta }^4-159540 \tilde{\beta }^3+209091 \tilde{\beta }^2-121238 \tilde{\beta }+240 \gamma  \left(225 \tilde{\beta }^4-717 \tilde{\beta }^3+879 \tilde{\beta}^2-455 \tilde{\beta }+110\right)+21056}{5760 \pi } \, , \\
    q_2^R &= -\frac{225 \tilde{\beta }^4-717 \tilde{\beta }^3+879 \tilde{\beta }^2-455 \tilde{\beta }+110}{24 \pi } \, , \\
    q_3^R &= -\frac{1017 \tilde{\beta }^4-3258 \tilde{\beta }^3+3948 \tilde{\beta }^2-2078 \tilde{\beta }+419}{96 \pi } \, , \\
    q_4^R &= -\frac{59931 \tilde{\beta }^4-187614 \tilde{\beta }^3+208836 \tilde{\beta }^2-95546 \tilde{\beta }+60 \gamma  \left(1017 \tilde{\beta }^4-3258 \tilde{\beta }^3+3948 \tilde{\beta}^2-2078 \tilde{\beta }+419\right)+15713}{2880 \pi } \, , \\
    q_5^R &= \frac{115002 \tilde{\beta }^4-357768 \tilde{\beta }^3+395655 \tilde{\beta }^2-181390 \tilde{\beta }+120 \gamma  \left(1017 \tilde{\beta }^4-3258 \tilde{\beta }^3+3948 \tilde{\beta }^2-2078 \tilde{\beta }+419\right)+28650}{5760 \pi } \, , \\
    %
    q_6^R &= \frac{532170 \tilde{\beta }^4-1687470 \tilde{\beta }^3+1937719 \tilde{\beta }^2-914459 \tilde{\beta }-10 \gamma  \left(4860 \tilde{\beta }^4-17460 \tilde{\beta }^3+22017\tilde{\beta }^2-9702 \tilde{\beta }+2776\right)}{57600 \pi } \nonumber \\
    &\quad + \frac{18803}{6400\pi}\, ,
\end{align}
\end{subequations}
for $F_R$. Higher orders can be derived systematically with the provided formulas. The logarithms in \eqref{eq:wc} are responsible for the non-polynomial structure of the inverse propagator discussed in the main text in the context of the unitarity mechanism. The coefficient $q_2^C$ ($q_2^R$) is directly related to $A_h$ ($\tilde A_h$), and the gauge dependence agrees with the one found in~\cite{Bonanno:2021squ}.

\section{Propagator of Faddeev-Popov ghosts}

In this appendix, we present the propagators and spectral functions of the transverse and longitudinal mode of the Faddeev-Popov ghost for the gauge choice $\beta=-1$. The propagators are shown in \autoref{fig:Gc}. As a consequence of the fact that the two respective anomalous dimensions are qualitatively and quantitatively similar, see \autoref{fig:eta_comparison}, the two propagators are also similar.

The two spectral functions are shown in \autoref{fig:rhoc}. In our gauge choice, they are both negative for all spectral values. This is, however, a gauge-dependent statement. For gauge choices $\beta>1.1$ ($\beta>0.68$), the transverse (longitudinal) spectral function is positive for small spectral values.

\begin{figure}[!h]
\centering
\includegraphics[width=0.49\linewidth]{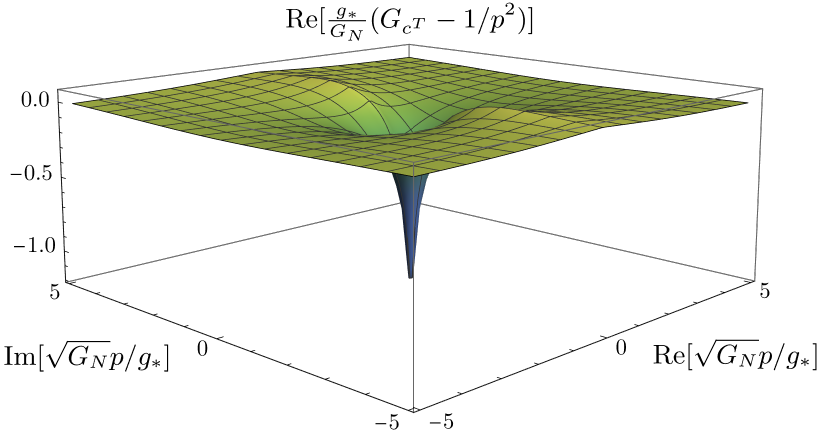} \hfill \includegraphics[width=0.49\linewidth]{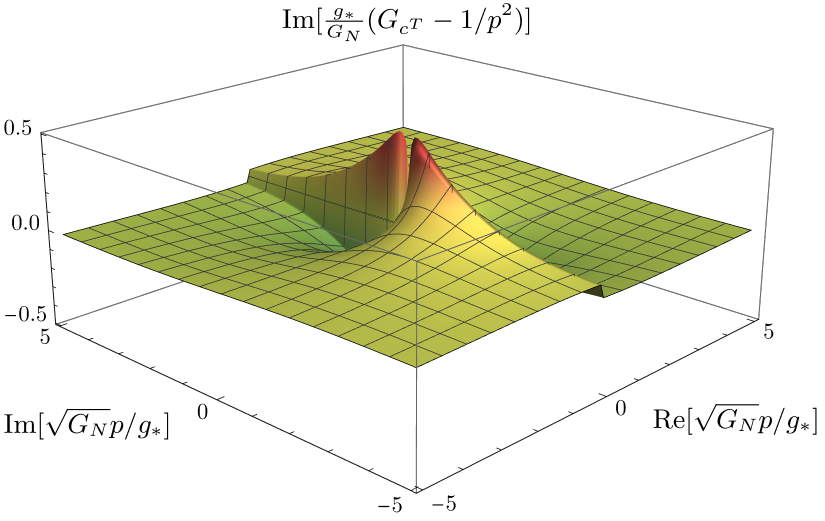} \\
\includegraphics[width=0.49\linewidth]{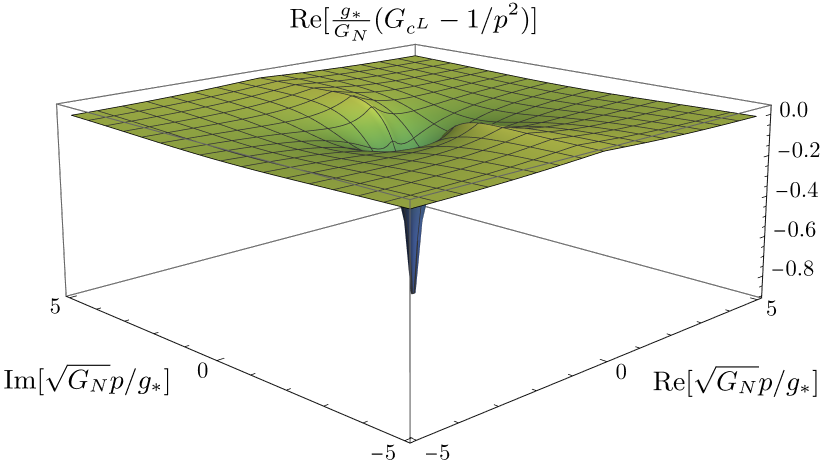} \hfill \includegraphics[width=0.49\linewidth]{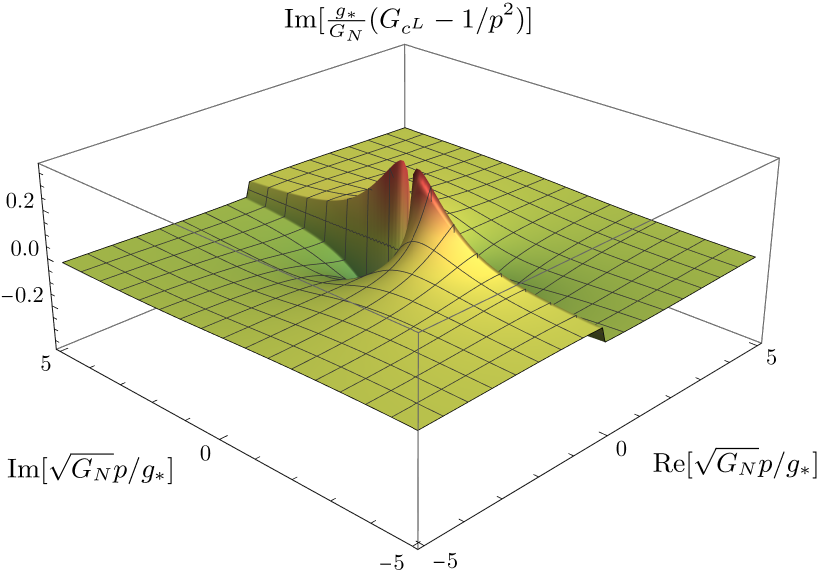}
\caption{\label{fig:Gc}Real and imaginary part of the transverse (upper panel) and longitudinal (lower panel) ghost propagator function $G_{c^{T,L}}$ in the complex plane in Planck units, as a function of momentum in Planck units. The massless pole at $p^2=0$ has been subtracted, and we have used $g_\ast=1/2$.}
\end{figure}

\begin{figure}[!t]
\centering
\includegraphics[width=0.49\linewidth]{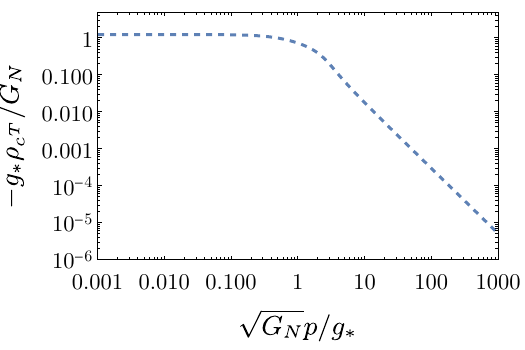} \hfill \includegraphics[width=0.49\linewidth]{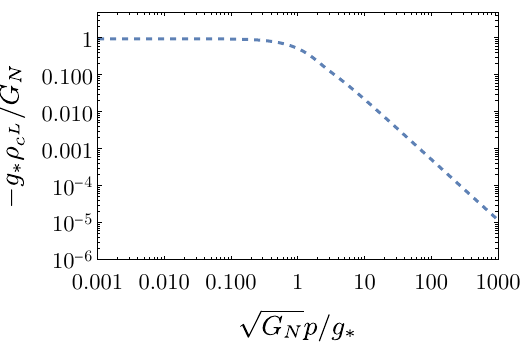}
\caption{\label{fig:rhoc}Spectral function $\rho_{c^{t,L}}$ of the transverse and longitudinal parts of the ghost in Planck units, as a function of momentum in Planck units. They are both negative on the entire positive real axis, indicated by the dashing. We have used $g_\ast=1/2$.}
\end{figure}

\clearpage

\end{widetext}

\end{document}